\documentclass[%
reprint,
superscriptaddress,
amsmath,amssymb,
aps,
pra,
]{revtex4-2}

\usepackage{graphicx}
\usepackage{dcolumn}
\usepackage{bm}
\usepackage{color}
\usepackage{setspace} 
\usepackage{comment}

\begin{document}

\preprint{APS/123-QED}

\title{A combined quantum-classical method applied to material design: optimization and discovery of photochromic materials for photopharmacology applications}

\author{Qi Gao}
 \email{caoch@user.keio.ac.jp}
\affiliation{Mitsubishi Chemical Corporation, Science \& Innovation Center, 1000, Kamoshida-cho, Aoba-ku, Yokohama 227-8502, Japan}
\affiliation{Quantum Computing Center, Keio University, Hiyoshi 3-14-1, Kohoku, Yokohama 223-8522, Japan}
\author{Michihiko Sugawara}
\affiliation{Quantum Computing Center, Keio University, Hiyoshi 3-14-1, Kohoku, Yokohama 223-8522, Japan}
\author{Paul D. Nation}
\affiliation{IBM Quantum, Yorktown Heights, New York 10598, USA}
\author{Takao Kobayashi}
\affiliation{Mitsubishi Chemical Corporation, Science \& Innovation Center, 1000, Kamoshida-cho, Aoba-ku, Yokohama 227-8502, Japan}
\affiliation{Quantum Computing Center, Keio University, Hiyoshi 3-14-1, Kohoku, Yokohama 223-8522, Japan}

\author{Yu-ya Ohnishi}
\affiliation{Quantum Computing Center, Keio University, Hiyoshi 3-14-1, Kohoku, Yokohama 223-8522, Japan}
\affiliation{Materials Informatics Initiative, RD technology and digital transformation center, JSR Corporation, 3-103-9, Tonomachi, Kawasaki-ku, Kawasaki, Kanagawa, 210-0821, Japan}
\author{Hiroyuki Tezuka}
\affiliation{Quantum Computing Center, Keio University, Hiyoshi 3-14-1, Kohoku, Yokohama 223-8522, Japan}
\affiliation{Advanced Research Laboratory, Technology Infrastructure Center, Technology Platform, Sony Group Corporation, 1-7-1 Konan, Minato-ku, Tokyo, 108-0075, Japan}
\author{Naoki Yamamoto}
 \email{yamamoto@appi.keio.ac.jp}
\affiliation{Quantum Computing Center, Keio University, Hiyoshi 3-14-1, Kohoku, Yokohama 223-8522, Japan}

\date{\today}

\begin{abstract}
Integration of quantum chemistry simulations, machine learning techniques, and optimization calculations is expected to accelerate material discovery by making large chemical spaces amenable to computational study; a challenging task for classical computers. 
In this work, we develop a combined quantum-classical computing scheme involving the computational-basis Variational Quantum Deflation (cVQD) method for calculating excited states of a general classical Hamiltonian, such as Ising Hamiltonian. We apply this scheme to the practical use case of generating photochromic diarylethene (DAE) derivatives for photopharmacology applications.
Using a data set of 384 DAE derivatives quantum chemistry calculation results, we show that a factorization-machine-based model can construct an Ising Hamiltonian to accurately predict the wavelength of maximum absorbance of the derivatives, $\lambda_{\rm max}$, for a larger set of 4096 DAE derivatives. 
A 12-qubit cVQD calculation for the constructed Ising Hamiltonian provides the ground and first four excited states corresponding to five DAE candidates possessing large $\lambda_{\rm max}$. 
On a quantum simulator, results are found to be in excellent agreement with those obtained by an exact eigensolver. 
Utilizing error suppression and mitigation techniques, cVQD on a real quantum device produces results with accuracy comparable to the ideal calculations on a simulator. 
Finally, we show that quantum chemistry calculations for the five DAE candidates provides a path to achieving large $\lambda_{\rm max}$ and oscillator strengths by molecular engineering of DAE derivatives. 
These findings pave the way for future work on applying hybrid quantum-classical approaches to large system optimization and the discovery of novel materials.   
\end{abstract}

\maketitle


\section{\label{sec:level1}Introduction}

Unprecedented improvements in the cost-effectiveness ratio of computers, together with improved computational techniques, enables quantum chemistry calculations to be widely applied in material design for finding novel molecules that have desirable properties {\color{blue}\cite{Pyzer-k_2015,Shoichet-B_2004,Hachmann-J_2014}}. On the other hand, since the number of candidate molecules is numerous, even utilizing supercomputing resources, it is time consuming to screen all molecules with $ab$ $initio$ calculations. Over the last 10 years, the use of machine learning and optimization methods to search for optimal molecular candidates in large chemical spaces has greatly accelerated {\color{blue}\cite{Saal-J_2020}} as the computational cost of these methods is several orders of magnitude lower than corresponding $ab$ $initio$ methods.
  
However, although using machine learning and optimization methods for material design shows great promise, there are fundamental problems associated with these methods that will need to be overcome before mainstream usage. One issue is related to the size and quality of data sets used for training machine learning models derived from quantum chemistry calculations. Due to the chemical diversity in standard data sets, the prediction of molecular properties from a data set trained by a machine learning model derived from differing data sources is challenging. Specifically, it has been shown that some under represented functional groups in the golden standard quantum chemistry-based QM9 data set {\color{blue}\cite{Ramakrishnan-R_2014}} is the primary source of outliers in property prediction {\color{blue}\cite{Glavatskikh-M_2019}}. Increasing the training data set size may improve the ability of machine learning models to predict molecular properties, however, such improvements come at the cost of requiring significantly more computational resources.

Another issue relates to the difficulties inherent in searching a large chemical space with discrete optimization methods. For example, while the number of potential drug-like molecules is estimated to be between $10^{23}$ and $10^{60}$, the number of all the synthesized molecules is $\sim 10^{8}$ {\color{blue}\cite{Polishchuk-P_2013}}. Since this small synthesized chemistry space is not a faithful representation of the full molecular space,  finding an optimal target from this reduced data set is a difficult task. Moreover, since brute force search is not tractable for large data sets, efficient heuristic algorithms, such as simulated annealing {\color{blue}\cite{Kirkpatrick-S_1983}} and Genetic algorithms {\color{blue}\cite{Mitchell-M_1998}}, are needed to find the solution in a reasonable amount of time. Work looking at converting a discrete optimization to a gradient-based multidimensional continuous optimization also seems promising {\color{blue}\cite{Bombarelli_2018}}.   

Quantum computing offers one potential route to remarkably speed up computational approaches for material design relative to classical computers. 
For some problems with computational time that scales exponentially with system size on classical computers, the corresponding time may increase only polynomially using quantum algorithms {\color{blue}\cite{Feynman-R_1985,Aspuru-Guzik_2005}}. 
Some recent studies also show that heuristic quantum optimization algorithms, such as the Variational Quantum Eigensolver (VQE) {\color{blue}\cite{farhi_2014}} and Quantum Approximate Optimization Algorithm (QAOA) {\color{blue}\cite{peruzzo_2014}}, may have better polynomial-time computational cost than that of the best classical heuristic methods {\color{blue}\cite{Kempe-J_2010,Egger-D_2021}}. 
A number of benchmark studies of applying quantum computing to the calculation of chemical reaction energy profiles {\color{blue}\cite{Gao-1_2021,Rice-J_2021}}, excited states energy spectra {\color{blue}\cite{Gao_2021,Ibe-Y_2021}} and searching lead molecules {\color{blue}\cite{Gao-2_2023}} demonstrate the potential of using quantum computing in new materials discovery.

\begin{figure}
\includegraphics[scale=0.12]{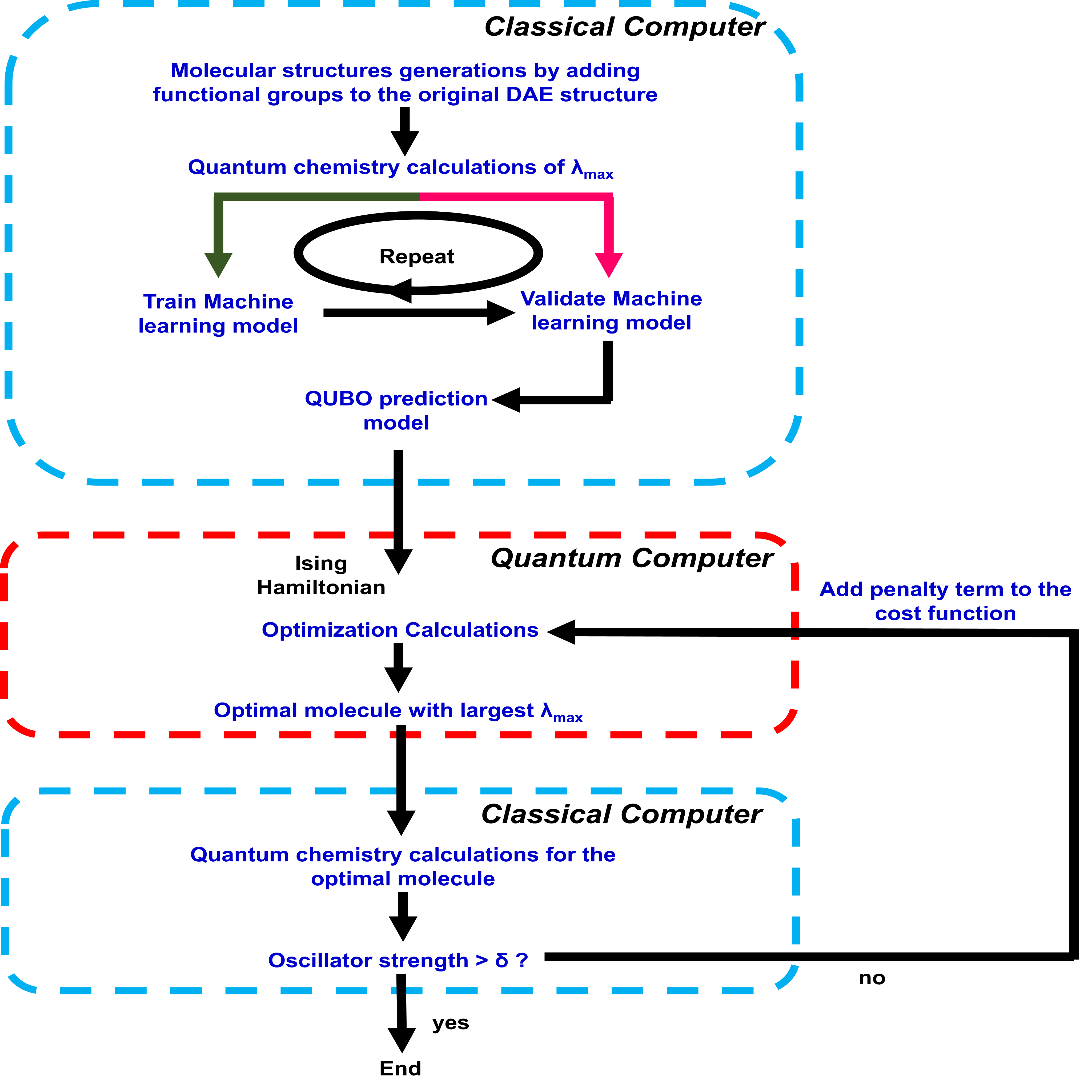}
\caption{\label{fig:fig1}
Flowchart of the hybrid quantum-classical method for identifying the optimal DAE derivatives with large $\lambda_{\rm max}$ and Osc.}
\end{figure}

In this work, we extend our previous quantum-classical computing scheme {\color{blue}\cite{Gao-2_2023}} to search for optimal materials with a set of required properties. The scheme uses quantum chemistry and machine learning on classical machinery followed by optimization on a quantum device as part of a workflow for materials discovery. 
Figure \ref{fig:fig1} shows the workflow typical for searching optimal prothochromic molecules, illustrating the concept of the proposed molecular design method. 
In detail, we first perform quantum chemistry calculations on a classical computer to find the target properties for a set of candidate molecules. 
We then use these results as a data set to construct an Ising Hamiltonian via machine learning for predicting target properties of the system. 
In particular, we perform a optimization computation on a quantum device, together with quantum chemistry calculations on classical machinery, to produce the ground and excited states of the constructed Ising Hamiltonian; 
these states serve as a set of candidate molecules that may satisfy required properties from which we can select the best molecule for the practical use. 

Optimization methods like QAOA and VQE are only used to calculate the ground state of a Hamiltonian, and thus these methods are limited to searching only one solution from a pool of candidates molecules, and are difficult to provide enough information to guide the design of novel molecules. 
For the purpose of searching excited states, in addition to the ground state, we employ a Variational Quantum Deflation (VQD) {\color{blue}\cite{Higgott-O_2019}} technique. This method computes the overlap between the ground state and the current state and takes advantage of the fact that if the Hamiltonian is limited to a classical one that is diagonal in the computational-basis (such as the Ising Hamiltonian), both the ground and excited states are the computational-basis states.  Accordingly, the process of calculating the overlap is reduced to computing the ratio of hitting the ground state on the computational-basis measurement result. This can drastically reduce the technical noise induced into VQD. To emphasize this advantage, we term this special type of VQD as the computational-basis VQD or cVQD.   Note that the SWAP test algorithm can be taken for the purpose of diagonalization, but at the expense of an increase in the introduced noise.

\begin{figure}
\includegraphics[scale=0.05]{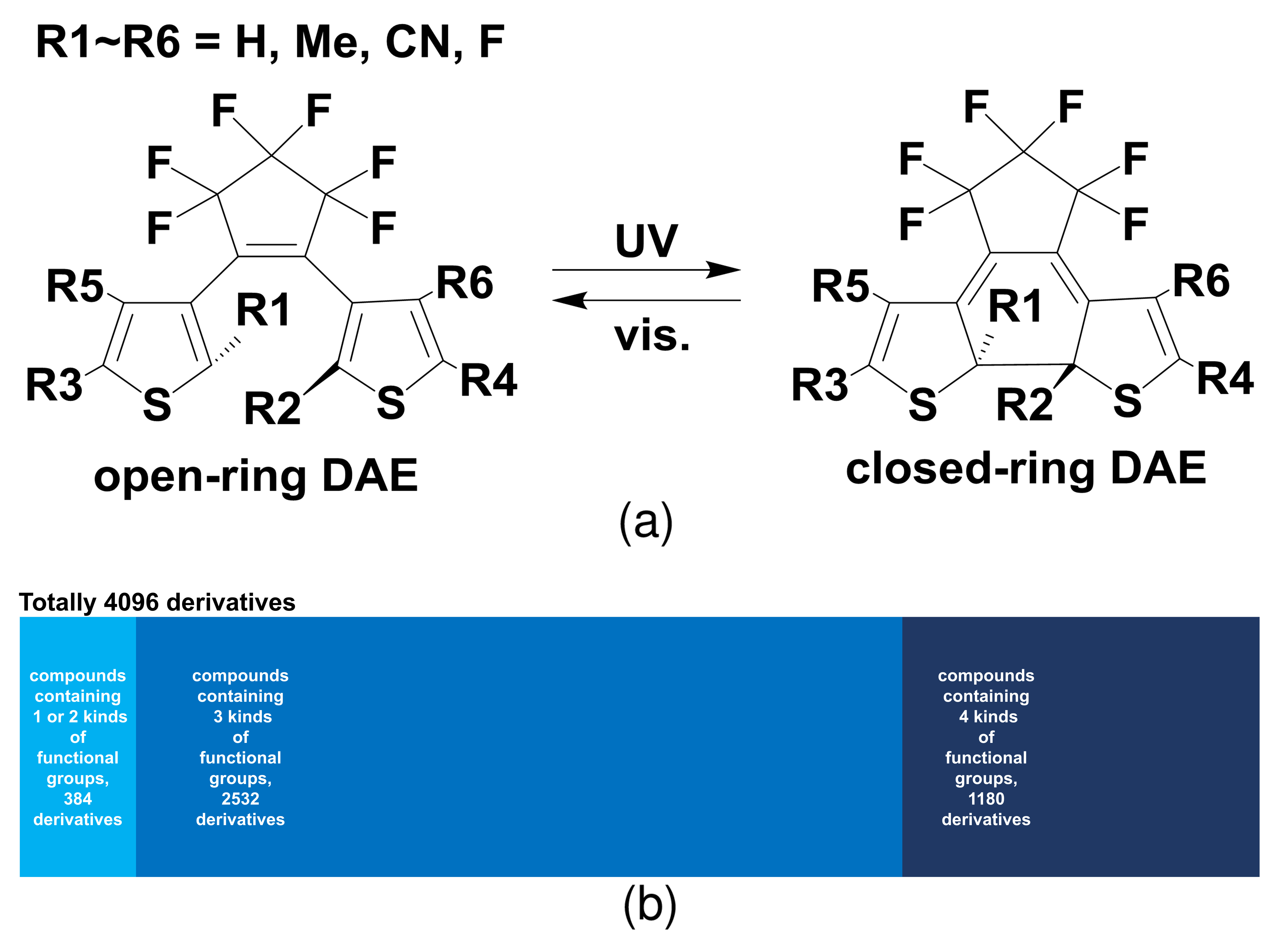}
\caption{\label{fig:fig2}
(a) Reversible photoisomerization of the most common DAE derivative and the numbering of 6 positions in thiophene rings used to generate various of DAE derivatives by bonding with funtional groups of -H, -Me, -CN, and -F. 
(b) Totally 4096 derivatives are classified as compounds containing 1 or 2, 3 and 4 kinds of functional groups.}
\end{figure}

We illustrate the ability of the combined quantum-classical method by taking a first step toward a go-beyond-concept industrial relevant use case of designing diarylethene (DAE) derivatives for photopharmacology application. Photopharmacology proposed by Feringa, a 2016 Nobel laureate in chemistry {\color{blue}\cite{Velema-W_2014}}, is an emerging technology in medicine which activates and deactivates photo-switchable molecules with light for target drug delivery to prevent side effects {\color{blue}\cite{Edwards_2000,Malhotra_2003}} and exposure to the environment of antibiotics {\color{blue}\cite{Carlet_2011,Martinez_2008}}. Among the prothochromic switch materials, DAE derivatives are particularly attractive compounds for use in photopharmacology due to their outstanding photochromic performance and thermal stability {\color{blue}\cite{Kudernac-T_2013,Irie-M_2014}}. The most common DAE derivative shown in Fig. \ref{fig:fig2} has reversible transformation between open-ring and closed-ring forms, where the open-ring and closed-ring forms have UV and Visible absorption spectra, respectively. To decrease the damage to tissues from light in photopharmacology applications, a key feature of designing DAE derivative is to make the open-ring form have a long wavelength of absorption light required to cause the photoisomerization.

The first step in designing DAE derivatives is generating a pool of 4096 DAE derivative candidates. 
We show that, by using the quantum chemistry calculation results of 384 DAE derivative as a training data set, the factorization-machine (FM)-based model can construct an Ising Hamiltonian which has 4096 states corresponding to 4096 DAE derivative; every state energy is represented via the wavelength $\lambda_{\rm max}$ in the absorption spectrum, such that the absorbance takes the maximum value of the corresponding DAE derivatives. 
In this way, the machine learning will accurately predict $\lambda_{\rm max}$ property in a much more rapid way, compared with quantum chemistry calculations.  

In the experiment, we performed VQE and QAOA methods to obtain the ground state; then we apply VQD and cVQD methods to obtain the first four excited state of the Ising Hamiltonian on a quantum simulator. 
We found that VQE outperforms the QAOA method in terms of both calculation accuracy and circuit depth. We also show that, in the absence of sampling error from quantum simulator, cVQD is capable of yielding markedly more accurate excited states than those from VQD. This is because the formalism in cVQD relies on the fact that the excited state of the Ising Hamiltonian is one of the computational basis, whereas VQD is originally developed for calculating the excited states of a quantum chemistry Hamiltonian.

After showing the ability of cVQD to accurately calculate the excited states of an Ising Hamiltonian, we then performed the same calculations on a quantum processor. We show that, although the noise from device leads to a decrease of calculation accuracy, cVQD can successfully produce accurate results for all the excited states. We also show that a rather accurate, e.g., better than a factor of 5, prediction of excited state can by achieved by using readout error mitigation and reducing decoherence noise via dynamical-decoupling error suppression.

Finally, we compute the oscillator strength (Osc) of 5 DAE derivatives corresponding to the ground and first four excited states of the Ising Hamiltonian with quantum chemistry methods on a classical computer. 
We also analysed the molecular orbitals contribution to the excitation. These results guide the design of new DAE deviates with large $\lambda_{\rm max}$ and Osc which are prerequisites for photopharmacology applications. 
Broadly speaking, these results show that the combination of quantum chemistry (on a classical computer), machine learning, and quantum computation could provide chemists with new tools in their arsenal for material discovery.

\section{Methods}

\subsection{Computational Methodology for discovery of photochromic materials}

Figure \ref{fig:fig1} shows the reversible photoisomerization of the most common DAE derivative (i.e. a cyclic 1,2-ethenedienyl bridge connecting two thiophene rings) between open-ring and closed-ring forms by the absorption of electromagnetic radiation, where the open-ring and closed-ring forms have UV and visible absorption spectra, respectively. Additional functional groups to the 6 positions of two thiophene rings can shift the absorption band of the open-ring form toward visible region, which is essential for the photopharmacology application. In this study, we identify open-ring candidates with long wavelength absorption and large Osc for the use as photo switch material through combinatorial optimization from 4096 DAE derivatives which are generated using -H, -Me, -CN and -F functional groups. 
By setting -H, -Me, -CN and -F as bit values of 10, 11, 00 and 01 respectively, the combinatorial optimization problems to find DAE derivatives which has required properties can be viewed as a problem for searching the corresponding bitstrings represented by 12 qubits on a quantum computer. Here, it is important to bear in mind that, since these 4 groups have different donor and acceptor strengths, the calculation results can also provide a path to tune the energy and probability of absorption in DAE molecules.

We here again describe the entire procedure for identifying DAE candidates with long wavelength absorption and large Osc, illustrated in Fig.~\ref{fig:fig1}. 
Once a number of DAE derivatives are chosen, the wavelength of maximum absorbance of the derivatives, $\lambda_{\rm max}$, is obtained from quantum chemistry calculations on a classical computer. 
With the calculation results, two data sets called training data set and test data set are prepared. 
The training data set is used to train a machine learning model to predict $\lambda_{\rm max}$. The test data set is used to validate the performance of the trained model by comparing the predicted values with those in the test data set. 
If the prediction accuracy of the trained model cannot achieve a desired threshold, further quantum chemistry calculation results are added to the training data set for training a better prediction model. 
The prediction model is repeated until the desired accuracy is achieved. 
With the improved model, an Ising Hamiltonian is constructed and is used to perform optimization on a quantum computer to find the DAE candidate which has the largest $\lambda_{\rm max}$. 
The Osc of the candidate are then obtained from quantum chemistry calculation on a classical computer. 
If Osc is below a chosen threshold, an "overlap" term is added to the cost function and additional optimization on a quantum computer is performed to find the DAE candidate which has the 2nd largest $\lambda_{\rm max}$. 
The procedure of the Osc calculation on classical computer and optimization on quantum computer is repeated until a DAE candidate with large $\lambda_{\rm max}$ and Osc is found. The following sections describe the details of the calculation in each stage.


\subsection{Construction of the Prediction Model}

All the initial structures of 4096 DAE derivative are generated using rdkit.Chem.rdChemReactions module in the RDkit Software {\color{blue}\cite{Landrum-G_2010}}. The minimum energy conformer of every DAE derivative is determined from the Universal force field {\color{blue}\cite{Rappe-A_1992}} in the same software. The determined conformer is used as the initial structure to perform ground state (S0) geometry optimization at the $\omega$B97X-D/def2-SVP {\color{blue}\cite{Chai-J_2008}} level with the density fitting (DF) approximation. The $\lambda_{\rm max}$ (i.e. the first singlet excited state (S1) absorption wavelength) and Osc (between S0 and S1 states) at the S0 optimized structures of 4096 DAE detivatives were calculated using the TD-$\omega$B97X-D/def2-SVP with the DF approximation. The (TD-)DFT {\color{blue}\cite{yanai_2004}} calculations were performed by Gaussian16 {\color{blue}\cite{frisch_2009}}. 


Setting functional groups of -H, -Me, -CN and -F as bit values of 10, 11, 00 and 01, every DAE derivative, which has these function groups on the positions of R1 to R6 as shown in Fig. \ref{fig:fig1},  is transformed into a binary feature vector $\vec{x}^{(s)}$ comprised of 12 qubits. A data set is prepared with all the $\vec{x}^{(s)}$  and its corresponding $\lambda_{\rm max}$ values of $y^{(s)}$ as shown in Table 1. We firstly chose all the DAE derivatives in which the structure has only one kind of functional groups as the training data. We also randomly select 60 derivatives from the structure containing two kinds of functional groups and add them to the training data. 
The chosen training data are used to train a prediction model described by quadratic unconstrained binary optimization (QUBO) formulation {\color{blue}\cite{kochenberger_2014}}:
\begin{eqnarray}
    E(q) &=& {\omega_0 +} \sum_{1 \le i \le j \le 12} Q_{ij}q_iq_j 
\nonumber \\
     &=& {\omega_0 +} \sum_{i=1}^{12} Q_{ii}q_i + \sum_{i=1}^{11} \sum_{j=i+1}^{12} Q_{ij} q_i q_j, 
\label{eq:two}
\end{eqnarray}
where {$\omega_0$ is a constant value,} $q_i$ is {the $i$th binary variable of $\vec{x}$} which takes either 0 or 1, and the coefficients $Q_{ij}$ and $Q_{ii}$ have real values. 
The second equality holds because $q_i^2=q_i$. The quantities $Q_{ij}$ and $Q_{ii}$ are obtained through the learning process; in particular, we use the FM predictor {\color{blue}\cite{rendle_fm_2010, rendle_2012, rendle_pairwise_2010}} with the following model equation:
\begin{eqnarray}
   \hat{y}(\vec{x}) &=& {\omega_0 +} \sum_{i=1}^{12} w_i q_i + \sum_{i=1}^{11} \sum_{j=i+1}^{12} (\vec{v}_i\cdot \vec{v}_j)q_iq_j, 
\label{eq:two}
\\ 
   && (\vec{v}_i\cdot \vec{v}_j) = \sum_{f=1}^\kappa v_{if} v_{jf},  
\label{eq:three}
\end{eqnarray}
where $\kappa$, the dimension of the factorization, is set to 8 during the learning process. The goal is to determine $(\omega_0, w_i, \vec{v}_i)$ and accordingly $Q_{ii}=w_i$ and $Q_{ij}=\vec{v}_i\cdot \vec{v}_j$, by minimizing the cost function $\sum_s(y^{(s)}-\hat{y}(\vec{x}^{(s)}))^2$. Note that although the QUBO model can also be determined by the regression of $Q_{ij}$ and $Q_{ii}$ in Eq.~(1), the advantage of using the FM predictor is that it can reliably fit the parameters $w_i$, $\vec{v}_i$ and $\vec{v}_j$ with a small training data set and prevent the trained model from over-fitting.

The performance of the QUBO model is evaluated by checking the correlation (i.e. correlation coefficient denoted by $R$) between the $\lambda_{\rm max}$ values of the molecular structures in test data set predicted by QUBO model and quantum chemistry calculations. 
If $R$ is less than a threshold value of 0.85, additional randomly selected 60 DAE derivatives, which have two kinds of functional groups in the structure, are added to the training data to retrain the QUBO model. The procedure repeats by adding every 60 DAE derivatives to  the training data until $R$ achieves the value larger than the threshold value. 
After all the derivatives containing two different functional groups (i.e. the derivatives in the light blue region of Fig. \ref{fig:fig4}(b)) are added to the training data set, the derivatives which have three and four kinds of the functional groups (i.e. the derivatives in the blue and dark blue regions of Fig. \ref{fig:fig4}(b)) are also used as the training data.

\begin{figure}
\includegraphics[scale=0.11]{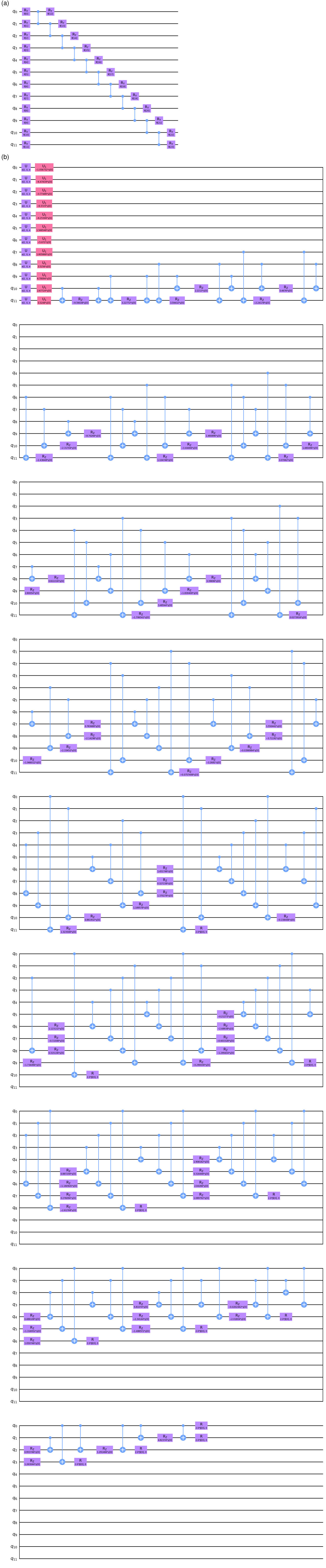}
\caption{\label{fig:fig3} 12-qubit circuits for (a) VQE with the $R_y$ Ans{\"a}tze with depth 1 and (b) QAOA with depth $p=1$.}
\end{figure}

\begin{table*}
\caption{\label{tab:table1}%
Dataset obtained from quantum chemistry calculations of DAE derivatives for training and validating the machine learning model. Every two bit values of 10, 11, 00 and 11 in the feature vector $\vec{x}^{(s)}$ correspond to the functional groups of -H, -Me, -CN and -F in the DAE derivatives.
}
\begin{tabular}{|p{1.0cm}|p{0.1cm}p{0.1cm}p{0.1cm}p{0.1cm}p{0.1cm}p{0.1cm}p{0.1cm}p{0.1cm}p{0.1cm}p{0.1cm}p{0.1cm}p{0.1cm}p{6cm}||p{1.0cm}|p{3cm}|}
\hline
\multicolumn{14}{|c||}{\textbf{feature vector $\vec{x}^{(s)}$ (functional groups of R1-R6)}} &
\multicolumn{2}{c|}{\textbf{target value $y^{(s)}$ ($\lambda_{\rm max}$)}}\\
\hline
\colrule
$\vec{x}^{(1)}$& 0 & 0 & 0 & 0 & 0 & 0 & 0 & 0 & 0 & 0 & 0 & 0 & \quad(CN-CN-CN-CN-CN-CN) &$y^{(1)}$ & 274.99\\
\hline
$\vec{x}^{(2)}$& 0 & 0 & 0 & 0 & 0 & 0 & 0 & 0 & 0 & 0 & 0 & 1 & \quad(CN-CN-CN-CN-CN-F) &$y^{(2)}$ & 282.63\\
\hline
$\vec{x}^{(3)}$& 0 & 0 & 0 & 0 & 0 & 0 & 0 & 0 & 0 & 0 & 1 & 0 & \quad(CN-CN-CN-CN-CN-H) &$y^{(3)}$ & 283.59\\
\hline
$\vec{x}^{(4)}$& 0 & 0 & 0 & 0 & 0 & 0 & 0 & 0 & 0 & 0 & 1 & 1 & \quad(CN-CN-CN-CN-CN-Me) &$y^{(4)}$ & 282.42\\
\hline
$\vdots$ & $\vdots$ & $\vdots$ & $\vdots$ & $\vdots$ & $\vdots$ & $\vdots$ & $\vdots$ & $\vdots$ & $\vdots$ & $\vdots$ & $\vdots$ & $\vdots$ & \quad(R1-R2-R3-R4-R5-R6) & $\vdots$ & $\vdots$\\
\hline
$\vec{x}^{(4094)}$& 1 & 1 & 1 & 1 & 1 & 1 & 1 & 1 & 1 & 1 & 0 & 1 & \quad(Me-Me-Me-Me-Me-F) &$y^{(4094)}$ & 282.52\\
\hline
$\vec{x}^{(4095)}$& 1 & 1 & 1 & 1 & 1 & 1 & 1 & 1 & 1 & 1 & 1 & 0 & \quad(Me-Me-Me-Me-Me-H) &$y^{(4095)}$ & 284.31\\
\hline
$\vec{x}^{(4096)}$& 1 & 1 & 1 & 1 & 1 & 1 & 1 & 1 & 1 & 1 & 1 & 1 & \quad(Me-Me-Me-Me-Me-Me) &$y^{(4096)}$ & 282.58\\
\hline
\end{tabular}
\end{table*}

\subsection{Optimizations on the quantum device}

\begin{figure}
\includegraphics[scale=0.08]{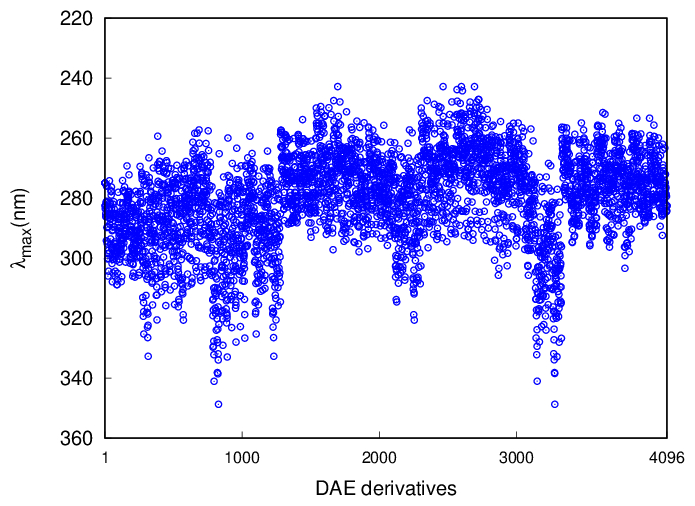}
\caption{\label{fig:fig4}$\lambda_{\rm max}$ of all the 4096 DAE derivatives from quantum chemistry calculations}
\end{figure}

To perform the optimization task on a quantum computer, the QUBO model $E(q)$ is converted into a spin-based Ising Hamiltonian {\color{blue}\cite{ising_1925, tanaka-book, tanahashi2019application}}:
\begin{equation}
   H = \sum_{i=1}^{12} h_i Z_i + \sum_{i=1}^{11}\sum_{j=i+1}^{12} J_{ij} Z_i Z_j, 
\label{eq:five}
\end{equation}
where $Z_i$ is the single-qubit Pauli $Z$ operator acting on the $i$th qubit. 
The optimization for calculating the ground states of Eq.~\eqref{eq:five} is performed using VQE and QAOA methods.

Ideally, VQE gives us the ground state energy of $H$ by minimizing the mean energy
\begin{equation}
E(\vec{\theta}) = \langle{\psi(\vec{\theta}) \vert H \vert \psi(\vec{\theta})}\rangle, 
\label{eq:six}
\end{equation}
with respect to the vector of parameters $\vec{\theta}$ of Ans{\"a}tze
\begin{equation}
\vert \psi(\vec{\theta})\rangle =  U(\vec{\theta})\vert \psi_0\rangle, 
\label{eq:seven}
\end{equation}
where $U(\vec{\theta})$ is the unitary operator of a parameterized quantum circuit and $|\psi_0\rangle$ is the initial state. 
In this work, $U(\vec{\theta})$ is set to the $R_y$ heuristic Ans{\"a}tze, which is constructed of only fixed CNOT gates and $R_y$ gates with angle parameters $\vec{\theta}$; the 12-qubit circuit featuring the $R_y$ Ans{\"a}tze has 11 nearest-neighbor CNOT gates and 24 optimization parameters $\vec{\theta}$ as shown in Fig.~3(a). 
The parameter $\vec{\theta}$ is updated so that the mean energy $E(\vec{\theta})$ decreases in each iteration. 

The parameterized Ans{\"a}tze state for QAOA is constructed from the Hamiltonian $H$ with single qubit Pauli $X$ operators as 
\begin{eqnarray}
     \vert \psi_p(\vec{\gamma},\vec{\beta)}\rangle 
         &=&  e^{-i\beta_p B} e^{-i\gamma_p H} \cdots 
               e^{-i\beta_1 B} e^{-i\gamma_1 H} \vert +\rangle^{\bigotimes 12}, 
\nonumber \\
   B &=& \sum_{i=1}^{12} X_i, ~
   \vert +\rangle = \frac{1}{\sqrt{2}}(\vert 0\rangle + \vert 1\rangle), 
\label{eq:eight}
\end{eqnarray}
where $p$ is an integer parameter representing the number of repetitions of the unitary operators. 
Like VQE, the parameters $\{(\gamma_i, \beta_i)\}_{i=1,\ldots,p}$ are updated so that the mean energy $E(\vec{\gamma},\vec{\beta}) = \langle{\psi_p(\vec{\gamma},\vec{\beta)} \vert H \vert \psi_p(\vec{\gamma},\vec{\beta)}}\rangle$ decreases in each iteration. 
An example of QAOA circuit with $p=1$ is shown in Fig.~3(b). 
The advantage of using the QAOA Ans{\"a}tze is that the adiabatic dynamics can be used to effectively find the ground state of the Hamiltonian to solve the combinatorical optimization problem.

All quantum optimization calculations were performed with the \texttt{statevector} simulator contained in the Aer module of Qiskit 0.37 {\color{blue}\cite{gadi_aleksandrowicz_2019}}, as well as on the 27-qubit IBM Quantum $ibm\_kawasaki$ quantum device. The \texttt{statevector} simulator uses linear algebra operations to compute the Ans{\"a}tze state and expectation values exactly. The parameter values in the VQE algorithm with the $R_y$ Ans{\"a}tze and QAOA were calculated using the Sequential Least Squares Programming (SLSQP) optimizer on the simulator. 
Since the computational cost of QAOA is rather demanding, only VQE calculations were performed on the quantum device with Simultaneous Perturbation Stochastic Approximation (SPSA) optimization. Twelve linearly connected qubits were selected to avoid the need for SWAP gates; see Appendix~\ref{app:device}.

\subsection{Computational-basis variational quantum deflation for the calculations of excited states of Ising Hamiltonian}

QAOA and VQE are only used to calculate ground state of Ising Hamiltonian and thus such methods are limited to search one solution from a pool of candidates molecules and are difficult to provide enough information to guide the design of novel molecules. For piratical use we need excited states for the Ising Hamiltonian in order to construct a set of candidate molecules, which, in our case, is a set of the DAE derivatives corresponding to different energy levels. 
For this purpose, we employ VQD {\color{blue}\cite{Higgott-O_2019}}, which uses a variational technique to compute the excited states for a general Hamiltonian. 
More specifically, the first excited state is calculated by minimizing the following modified cost function:
\begin{equation}
E(\vec{\theta}) = \langle{\psi(\vec{\theta}) \vert H \vert \psi(\vec{\theta})}\rangle + \beta_0\vert\langle{\psi(\vec{\theta}) \vert \psi_0}\rangle\vert^{2}, 
\label{eq:nine}
\end{equation}
where $\vert\psi_0\rangle$ is the reference ground state; that is, at first we need to prepare the ground state for using VQD. 
That is, VQD decreases the energy under the constraint that the current state $\vert\psi(\vec{\theta})\rangle$ is orthogonal to the reference ground state $\vert\psi_0\rangle$. 
The parameter ${\beta}_0$ represents the weight of the constraint. 
Ideally, $\vert\psi(\vec{\theta})\rangle$ converges to the first excited state.

cVQD is a special type of VQD, which focuses on a general classical Hamiltonian that is diagonal in the computational basis $\{|k\rangle\}$. 
The ground state is one of them, which we denote as $| k_0 \rangle$. 
Now, let us represent the current state in the computational-basis as follows: 
\[
   | \psi(\vec{\theta})\rangle = \sum_k \alpha_k(\vec{\theta}) | k \rangle. 
\]
By substituting this equation for Eq.~\eqref{eq:nine}, we have 
\begin{equation}
E(\vec{\theta}) 
= \langle{\psi(\vec{\theta}) \vert H \vert \psi(\vec{\theta})}\rangle 
+ \beta_0 |\alpha_{k_0}(\vec{\theta})|^2. 
\label{eq:ten}
\end{equation}
Minimization of Eq.~\eqref{eq:ten} with $\beta_0\gg 1$ will give a good approximated state of the first excited state. 
A clear advantage of cVQD is that the penalty term $|\alpha_{k_0}(\vec{\theta})|^2$ can be easily estimated on classical computer by using the ratio of hitting the ground state on the computational-basis measurement result. 
This is in stark contrast to VQD, where the calculation accuracy of $\vert\langle{\psi(\vec{\theta}) \vert \psi_0}\rangle\vert^{2}$ is largely influenced by the noise from the quantum device. 
Thus, cVQD is an efficient and robust algorithm for specifically calculating excited states of a classical Hamiltonian such as the Ising Hamiltonian. 
In Section {\color{blue}III-B}, we will show that, on a noisy quantum simulator, cVQD provides much more accurate results in searching the excited states than the standard VQD protocol that uses SWAP test.

\section{Results and Discussion}

\subsection{Quantum Chemistry and Machine Learning}

\begin{figure*}
\includegraphics[scale=0.3]{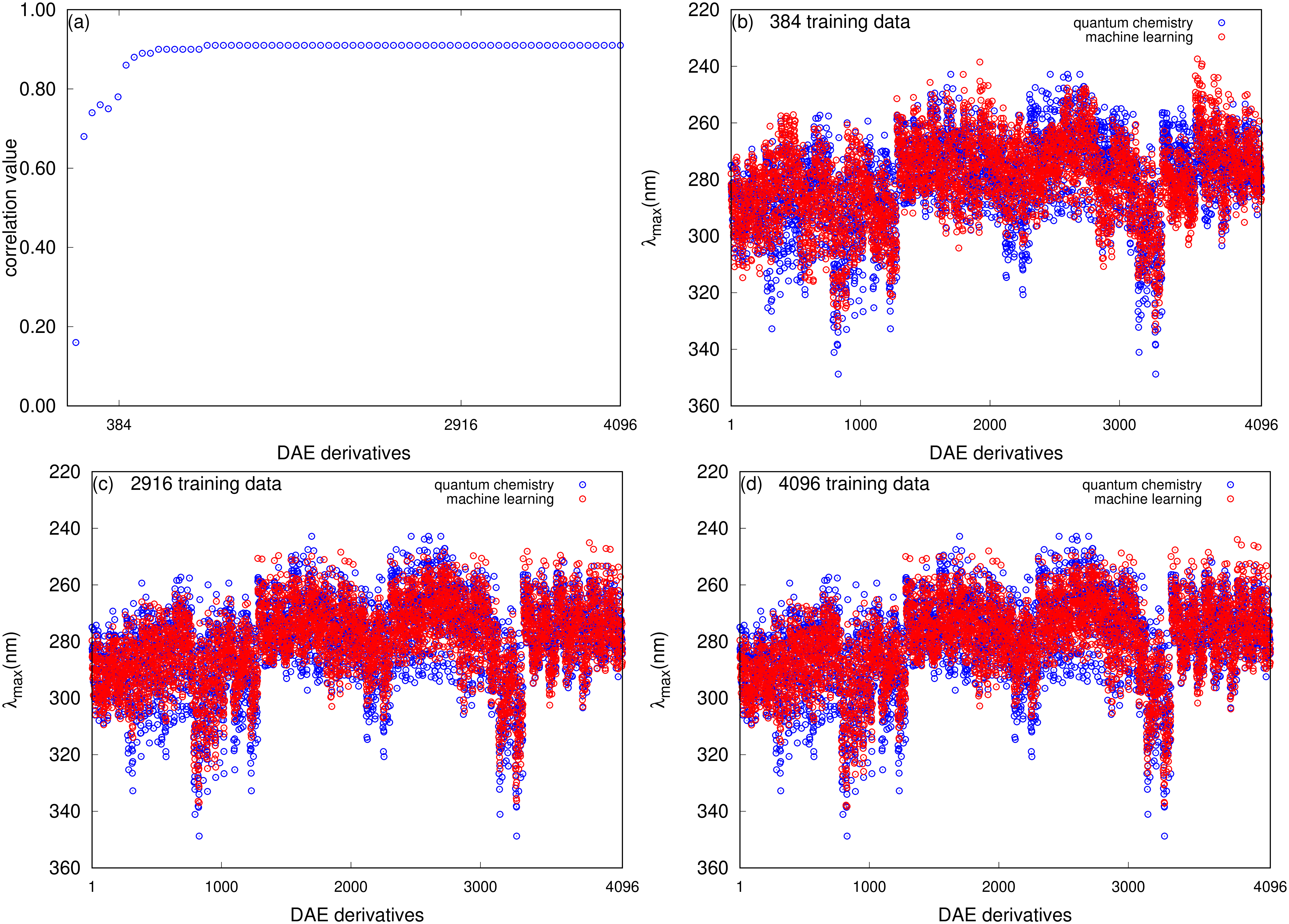}
\caption{\label{fig:fig5} Correlation values between $\lambda_{\rm max}$ obtained from the quantum chemistry calculations and QUBO models, as a function of a training data size, is shown in (a). 
Predicted $\lambda_{\rm max}$ for all the DAE derivatives by the quantum chemistry calculations and QUBO models, trained by the FM predictor with (b) 384, (c) 2916 and (d) 4096 DAE derivatives.}
\end{figure*}

Figure~\ref{fig:fig4} shows $\lambda_{\rm max}$ of all the 4096 DAE derivatives obtained from the quantum chemistry calculations. 
We found that, by changing the functional groups bonded to the thiophene rings, DAE derivatives exhibits $\lambda_{\rm max}$ in the range from 242.83\,nm to 348.75\,nm. 
Such wide range indicates that it may be sufficient to modify the absorption spectrum by adding functional groups to the DAE molecule. 
The DAE derivatives that showed the first and second largest $\lambda_{\rm max}$ among all the derivatives were CN-Me-CN-Me-H-Me and Me-CN-Me-CN-Me-H.
Functional groups of -Me which have strong donor strengths are found in one thiophene ring of these two derivatives. 
Similarly, functional groups of -CN which have strong acceptor strengths are found in the another thiophene ring. 
These results mean that $\lambda_{\rm max}$ may be sufficiently tunable by modifying the substituents of the thiophene ring moieties to modulate their donor and acceptor strengths.

The ability of the machine learning model obtained from the FM approach with different number of training data to accurately predict $\lambda_{\rm max}$ values is investigated by comparing the results predicted by the QUBO model with quantum chemistry calculations, as shown in Fig.~\ref{fig:fig5}. 
With the smallest training data set which has 64 DAE derivatives, the QUBO model fails to accurately predict $\lambda_{\rm max}$ values, i.e., the correlation of $R$ between $\lambda_{\rm max}$ values from the QUBO model and quantum chemistry calculation is 0.16. 
Increase of the training data set size from 64 to 436 derivatives significantly improves $R$ to 0.86, while further increase of the data set size from 436 to 1036 derivatives has only a minor improvement of $R$ by 0.05. 
With the size bigger than 1036 derivatives, the accuracy of the the QUBO model appears to level off. 
Comparison of $\lambda_{\rm max}$ predicted by the QUBO model with the quantum chemistry calculation values also gives evidence that the QUBO model slightly underestimates $\lambda_{\rm max}$ for the derivatives which have relatively small and large $\lambda_{\rm max}$ values. 
The underestimation of $\lambda_{\rm max}$ is possibly attributed to the imbalanced distribution of the training data set. 
In detail, $\lambda_{\rm max}$ of a large proportion of DAE derivatives takes values in the range 270-300\,nm. 
Thus, the training model spends most of its time on the large proportion of the data set and cannot learn enough from the DAE derivatives of which $\lambda_{\rm max}$ is smaller than 270\,nm or larger than 300\,nm. 
Nevertheless, since the trend of $\lambda_{\rm max}$ predicted by the QUBO model are in quantitative agreement with results of quantum chemistry calculations and $\lambda_{\rm max}$ of the suggested DAE derivative will be finally determined by quantum chemistry calculations, the performance of the QUBO model is considered to be good enough to be used for the purpose of finding the optimal DAE derivative.

Figures~\ref{fig:fig5}b-d show that the $\lambda_{\rm max}$ distribution obtained from the quantum chemistry calculations is similar to those from the QUBO model trained by using the dataset of DAE derivatives containing 1-2 kinds (Fig. \ref{fig:fig5}b), 1-3 kinds (Fig. \ref{fig:fig5}c) and 1-4 kinds (Fig. \ref{fig:fig5}d) of functional groups. 
These results suggest that the FM approach can learn $\lambda_{\rm max}$ values well, using the 384 DAE derivatives which have 1-2 kinds of functional groups in the structures. 
The good performance of the QUBO model trained by the small data set of 384 DAE derivatives is presumably due to that the $\lambda_{\rm max}$ values of most of all the DAE derivatives can be mainly described as a transition between two molecular orbitals localized on the two thiophene rings of DAE derivatives. 
Since the number of the DAE derivatives grows on the order of $n^6$ with $n$ the number of functional groups, even with $n=10$ it would be a daunting task to explore an optimal DAE derivative with desired $\lambda_{\rm max}$ and Osc using quantum chemistry approaches. 
Our results show the possibility that, with machine leaning approach, the computational cost of quantum chemistry calculations can be reduced to the order of $2^6n^2$. 

\begin{figure*}
\includegraphics[scale=0.3]{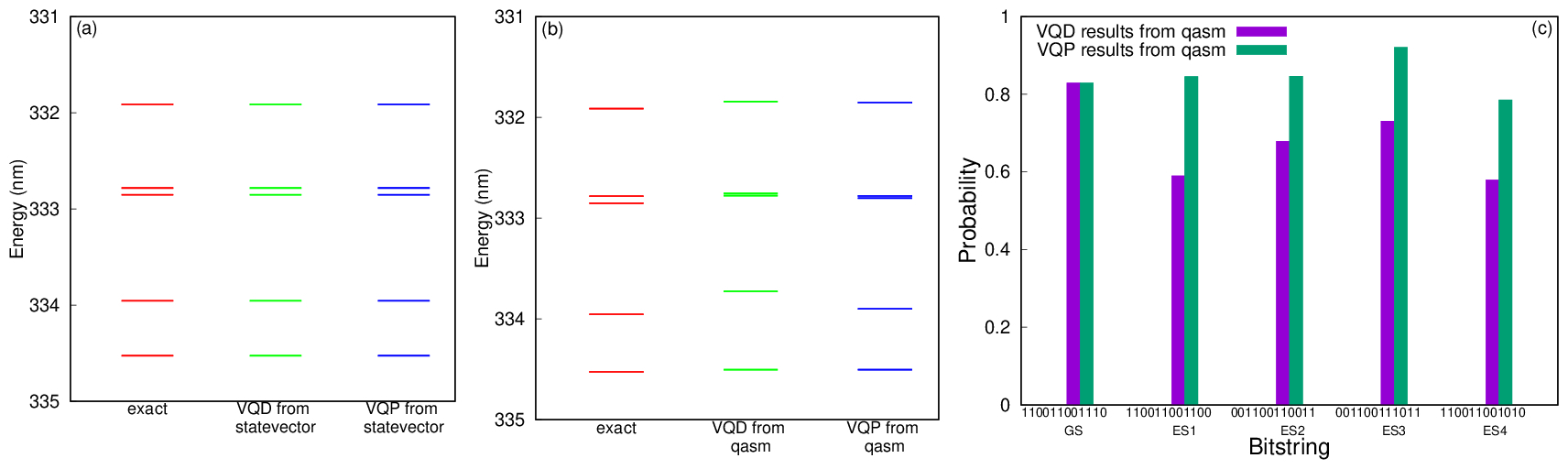}
\caption{\label{fig:fig6}Energies of the ground state (GS) and 1-4 excited states (ES1-ES4) of Ising Hamiltonian obtained from the exact diagonalization, VQD and cVQD calculations with (a) statevector and (b) qasm simulators. 
The highest probability bitstrings of the GS and ES1-ES4 obtained from VQD and cVQD calculations with qasm simulator is shown in (c).}
\end{figure*}

\subsection{Simulations with the \texttt{statevector} and \texttt{qasm} simulator}

\begin{table*}
\caption{\label{tab:table2}%
Ground state energies and corresponding bitstrings appearing with the highest probability, computed using various Ans{\"a}tze with the \texttt{statevector} simulator. In addition, Ans{\"a}tze properties comprising the number of CNOT gates and the number of optimization parameters (opt params) are shown. 
}
\begin{tabular}{|p{3cm}|p{1.5cm}|p{8cm}|p{1.5cm}|p{1.5cm}|p{1.5cm}|}
\hline
\multicolumn{1}{|c|}{\textbf{Method}} &
\multicolumn{3}{c|}{\textbf{Ground state}} &
\multicolumn{2}{c|}{\textbf{Ans{\"a}tze}} \\
\hline
\multicolumn{1}{|c|}{\textrm{}} &
\multicolumn{1}{|c|}{\textrm{energy}} &
\multicolumn{1}{|c|}{\textrm{bitstring (functional groups of R1-R6)}} &
\multicolumn{1}{|c|}{\textrm{probability}} &
\multicolumn{1}{|c|}{\textrm{CNOTs}} &
\multicolumn{1}{|c|}{\textrm{opt params}}\\
\hline
exact eigensolver & 334.527 & [110011001110] (Me-CN-Me-CN-Me-H) & 1.00 & - & -  \\
\hline
VQE & 334.527 & [110011001110] (Me-CN-Me-CN-Me-H) & 1.00 & 11 & 24 \\
\hline
QAOA $(p=1)$ & 313.827 & [110011001110] (Me-CN-Me-CN-Me-H) & 0.05 & 132 & 6 \\
\hline
QAOA $(p=2)$ & 323.982 & [110011001110] (Me-CN-Me-CN-Me-H) & 0.15 & 264 & 14 \\
\hline
QAOA $(p=3)$ & 332.051 & [110011001110] (Me-CN-Me-CN-Me-H) & 0.25 & 396 & 22 \\
\hline
\end{tabular}
\end{table*}

By converting the QUBO model to an Ising model \eqref{eq:five}, the problem of searching an optimal DAE derivative can be mapped to a quantum optimization problem whose ground state corresponds to the optimal DAE derivative. 
The ground state energy and its corresponding bit-strings calculated by the exact eigensolver (i.e., an exact diagonalization of the Ising model Hamiltonian), VQE and QAOA algorithms on the \texttt{statevector} simulator for the Ising models, are shown in Table~\ref{tab:table2}. 
VQE with a heuristic $R_y$ Ans{\"a}tze reproduces the ground state obtained from the exact eigensolver. 
The bit-string of (110011001110) which corresponds to the optimal DAE derivative has the probability of 1.00, meaning that this correct solution is always obtained. 
That is, the quantum optimization using VQE algorithm successfully found the exact ground state of the Ising model Hamiltonian. 
On the other hand, unlike VQE calculation, all three type of QAOAs with $p=1,2,3$ did not converge to the exact ground state; still, in every case of $p$, the most probable bit-strings of the converged state corresponds to the optimal DAE derivative. 
The converged energy of the QAOA calculations increases and its corresponding highest probability of the bit-string also increases, with increasing the value of $p$ in Eq.~\eqref{eq:eight}. 
The improvement with the increase of the value $p$ is presumably due to that the usage of a large value of $p$ gives an appreciably more accurate description of the time-evolution operator \cite{farhi_2014}. 
On the other hand, even with $p=3$, a relatively larger deviation of 2.476\,nm between the QAOA calculation result (331.051\,nm) and the exact value (334.527\,nm) is observed; moreover, the highest provability value of hitting the correct solution is 0.25, which is lower than that of VQE calculation by a factor of 4. 
Further increase of $p$ should improve the ability of the QAOA calculation to accurately converge to the ground state; however, such improvements add significant computational expense on both simulator and quantum devices. 
These results suggest that VQE outperforms QAOA in terms of both calculation accuracy and circuit depth.

Based on the higher accuracy and lower computational cost of ground state calculations using VQE than QAOA, we decide to use the heuristic $R_y$ Ans{\"a}tze to calculate 1st to 4th excited states of the Ising Hamilitonian with VQD and cVQD. 
The energies of the ground and the four excited states calculated by the exact eigensolver, VQD and cVQD algorithms on the \texttt{statevector} simulator are shown in Fig. \ref{fig:fig6}(a). 
The same results are predicted by 
VQD and cVQD for the energies of the ground state, due to the fact that both algorithms use the same VQE to calculate the zeroth approximator for the ground state. 
Although VQD and cVQD use different strategies to calculate excited states, both predict very similar energies for all the four excited states on \texttt{statevector} simulator. 
Moreover, both VQD and cVQD well 
reproduce the excited energies predicted by the exact eigensolver.
These results imply that both VQD and cVQD can accurately predict excited state energies of the Ising Hamiltonian on \texttt{statevector} simulator.

Using VQD and cVQD for calculating excited states of the Ising Hamiltonian on \texttt{statevector} simulator shows great promise, but 
there is a fundamental problem associated with performing VQD and cVQD on \texttt{qasm} and quantum device; 
namely, the fact that sampling and device noise influence the overlap term, which is necessary to search the excited states. 
Figure~\ref{fig:fig6}(b,c) compares all the calculated excited state energies of the Ising Hamiltonian and their corresponding highest probability bitstrings using VQD and cVQD on a \texttt{qasm} simulator with 81920 shots. 
The excited state energies obtained from the calculation using VQD is 0.085\,nm higher than the exact values, in contrast to that obtained by using cVQD which is only 0.038\,nm higher. 
These results indicate that the excited states computed 
from cVQD possesses more fidelity with the exact excited state than that obtained by using 
VQD.

More importantly, 
VQD leads to a maximum error of 0.23\,nm higher than the energies obtained with the exact eigensolver, whereas 
cVQD results in a maximum error that are only 0.06\,nm larger than the exact values. 
These results signify the effectiveness of cVQD to accurately calculate all the overlap terms which are needed to reliably calculate excited states. 
Figure~\ref{fig:fig6}(b,c) also indicates that the accuracy of the cVQD calculations may be invariant to changes in the level of excited state, because the exact states are used as the reference states in cVQD. 
Accordingly, although the bitstrings appearing with the highest probability for all the excited states are the same between the VQD and cVQD calculations, VQD yields probabilities an average of 0.16 lower than cVQD.

The number of CNOT gates in the circuit used for VQE and QAOA is shown in Table \ref{tab:table2}; this information shows which optimization algorithms are suitable for use on the $ibm\_kawasaki$ quantum device for the 12-qubit system calculation. 
Quantum circuits for VQE with the $R_y$ Ans{\"a}tze possess 11 CNOT gates, whereas 132 CNOT gates are required for the QAOA calculation with $p=1$. 
Furthermore, the number of CNOT gates linearly increases with the value of $p$ with the proportional coefficient 132. 
Note that, since the qubits of $R_y$ Ans{\"a}tze are connected in a line shape, this Ans{\"a}tze can be directly implemented on $ibm\_kawasaki$, without additional qubits for SWAP operations. 
On the other hand, since QAOA does not have such linear structure for connecting the qubits, overhead of the number of CNOT gates by a factor of 7-8 is needed for implementing this Ans{\"a}tze on a quantum device. 
Because the CNOT error rate largely influences on the computation result, VQE is expected to provide higher accuracy than those provided by QAOA. 
Thus, we decided to use $R_y$ Ans{\"a}tze to implement VQE and cVQD to obtain the ground and excited states, respectively, of Ising Hamiltonian on $ibm\_kawasaki$.

\begin{figure*}
\includegraphics[scale=0.3]{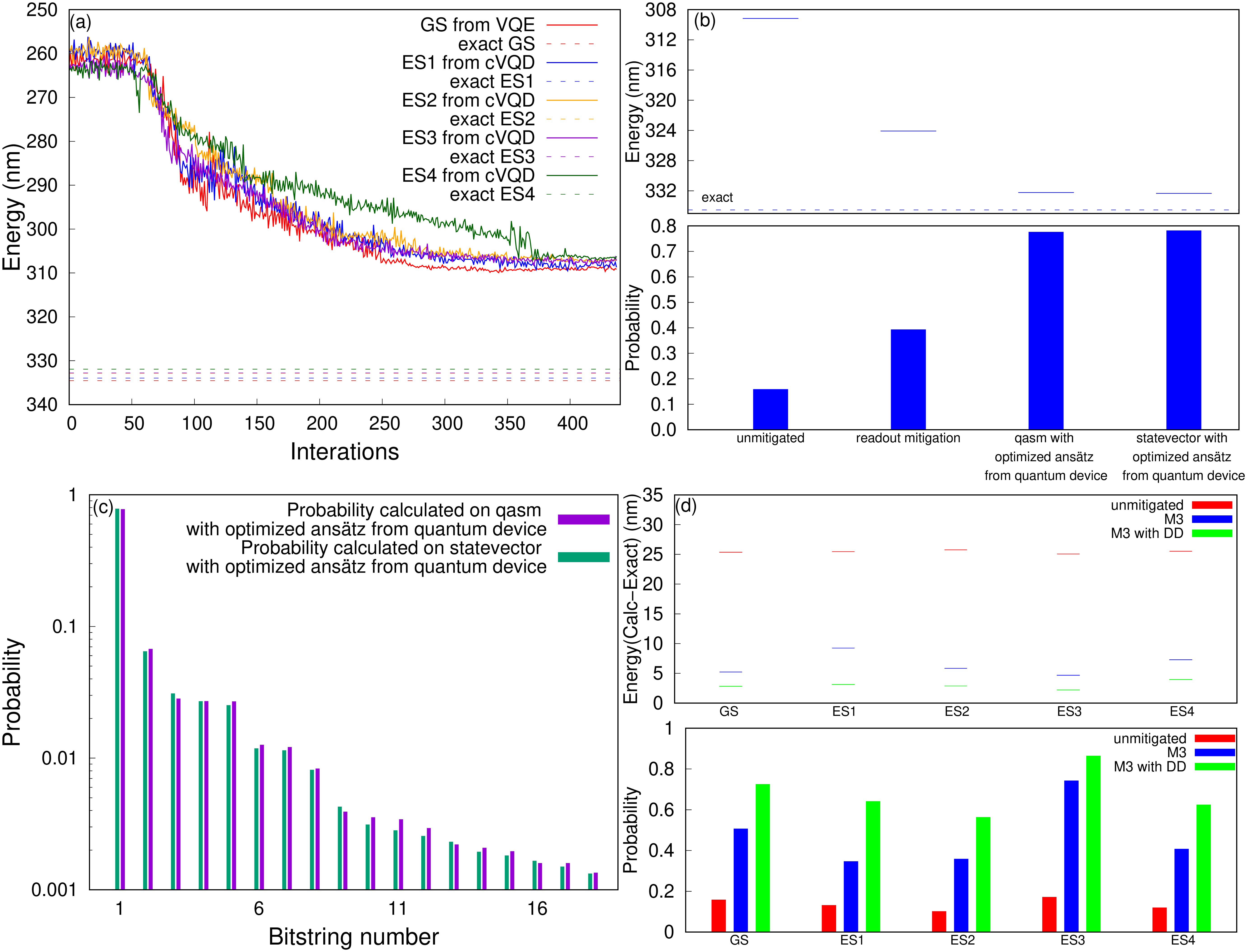}
\caption{\label{fig:fig7} The energy minimization process of VQE and cVQD with $ibm\_kawasaki$ is shown in (a). 
Energies ((b) upper panel) and the highest probability bitstring ((b) lower panel) of the ground states obtained from the $ibm\_kawasaki$, qasm and statevector simulations with $R_y$ Ans{\"a}tze. 
Optimized parameters for $R_y$ Ans{\"a}tze were determined from VQE calculations on $ibm\_kawasaki$. 
Bitstrings versus their occurring probabilities, for the states obtained from qasm and statevector simulators(c). 
The energies ((d) upper panel) and the highest probability bitstrings ((d) lower panel) of the GS and ES1-ES4 obtained from the unmitigated $ibm\_kawasaki$, $ibm\_kawasaki$ with M3 and M3/DD methods.}
\end{figure*}

\begin{figure*}
\includegraphics[scale=0.15]{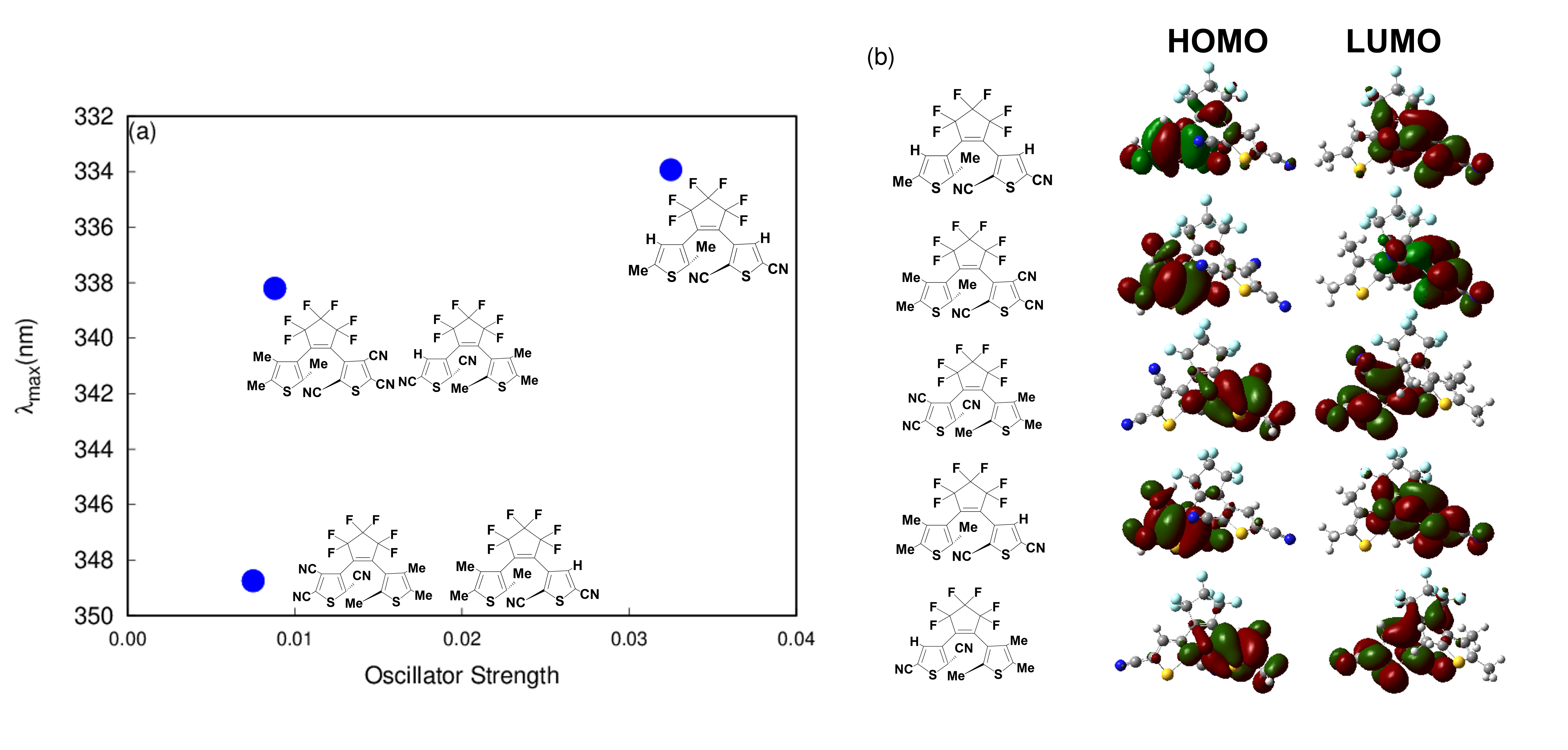}
\caption{\label{fig:fig8} 
(a) Quantum chemistry calculation results of $\lambda_{\rm max}$ as a function of oscillator strengths. 
(b) The HOMO, LUMO for the five DAE candidate derivatives suggested by cVQD calculations on $ibm\_kawasaki$.}
\end{figure*}

\subsection{Calculations on $ibm\_kawasaki$}%

Simulations using the VQE and cVQD algorithms with an Ising Hamiltonian and $R_y$ Ans{\"a}tze in Fig.~\ref{fig:fig3}(a) were performed on the 27-qubit $ibm\_kawasaki$ device \footnote{Note that, although the controlled-z gates in Fig.~\ref{fig:fig3}(a) can be executed in parallel, we choose to execute them in serial, with idle times populated with dynamical decoupling sequences as it yields higher fidelity results. This is likely due to the always-on-coupling of cross-resonance based devices, where the run-in-parallel gate operation can have larger error rates compared to isolated execution.} with 8192 shots. 
Detailed system information can be found in Appendix~\ref{app:device}. 
The energy values returned as a function of the number of iterations for each method are shown in Fig.~\ref{fig:fig7}(a). 
The same convergence trend is observed between energies calculated from VQE and cVQD on $ibm\_kawasaki$ toward the exact eigensolver solution, with VQE and cVQD calculations predicting energies several nm higher than the exact values. 
Specifically, the converged energies of the ground state and the first four excited states are found to be 25.36, 25.24, 25.67, 25.62 and 25.71 nm higher, respectively. 
We note that, even on the noisy present-day quantum device, the cVQD method is capable of accurately computing the energy level of the nearly degenerate 2nd and 3rd excited states. 
The results reveal that, through the use of the overlap term in Eq.~(\ref{eq:ten}), cVQD is capable of faithfully computing the excited states of the Ising Hamiltonian on a quantum device.

Extending the analysis to understand the influence of noise from the quantum device on the accuracy, we compare the ground state energies and the highest probability bit-string obtained by the use of exact eigensolver with those obtained from $ibm\_kawasaki$ with the readout error mitigation, \texttt{qasm} and \texttt{statevector} simulations with the parameter optimized $R_y$ Ans{\"a}tze. Note an efficient readour error mitigation technique called M3 (i.e. Matrix-free Measurement Mitigation) {\color{blue}\cite{Nation_2021}} is used to  approximately cancel measurement assignment error.
Figure~\ref{fig:fig7}(b) shows that VQE calculations without the error mitigation yield energies about $25.36$\,nm higher than the exact value. 
This difference is reduced to $\sim 10.45$\,nm using the readout error mitigation. 
Accordingly, the value of the highest probability of the bit-string is improved from $0.16$ to $0.39$. 
We also found that the energies from \texttt{qasm} simulations with optimized $R_y$ Ans{\"a}tze obtained using VQE on $ibm\_kawasaki$ has an error of $2.29$ nm to the exact value, and the corresponding highest probability of the bit-string is $0.78$. 
The error and the probability is almost the same as those obtained from \texttt{statevector} simulations. Moreover, we observe that the probability values for the remaining bit-strings from \texttt{qasm} simulations are also similar to those from \texttt{statevector} simulations, as shown in Fig. \ref{fig:fig7} (c). These results suggest that the readout error is the dominant noise source (64\% effect), with only 0.5\% due to finite-sampling limitations placed on measurements of single circuit executions (i.e. shots). The effect to the remaining error is considered from other known sources including depolarization, dephasing, and deviations of quantum gates from their
normal unitary operations (coherent errors).

Having identified the main noise sources on our device, we apply readout error mitigation to the resulting noisy distribution. 
Here we use the M3 method that corrects the probability with a matrix-free preconditioned iterative-solution. 
Moreover, we apply Dynamical Decoupling (DD) {\color{blue}\cite{Souza_2012}} to reduce decoherence by repetitively inserting CPMG sequences {\color{blue}\cite{Gill_1958}} into the idle periods between gates. 
Figure~\ref{fig:fig7}(d) shows that energies of the ground state and the first four excited states obtained from VQE and cVQD without error mitigation approaches are on average 25.44\,nm larger than the exact values. 
Using the readout error mitigation alone, the difference is improved to $\sim 6.45$\,nm. 
However, applying both the error suppression (i.e., DD) and the error mitigation yields the best results giving energies that are within $3$\,nm of the exact values. 
The corresponding improvement of the highest probability bit-strings for the ground and first four excited states is improved from $0.14$ to $0.68$ using DD and M3 together.

Overall, given the influence of the noise from quantum device to the calculation accuracy, VQE and the proposed cVQD methods give the highest probability bit-strings corresponding to the exact ground and excited states, and thus can be applied to finding the optimal DAE derivatives. 
Moreover, through the combined use of error suppression and readout error mitigation, the accuracy of the energy differences to the exact values can be improved by a factor of $\sim 8$ and the corresponding bit-string probabilities improved by $5x$.

\subsection{DAE Molecular design strategy for photopharmacology application }

$\lambda_{\rm max}$ and Osc of five candidate DAE derivatives obtained from VQE and cVQD on $ibm\_kawasaki$ were calculated at the TD-$\omega$B97X-D/def2-SVP level and are shown in Fig.~\ref{fig:fig8}(a). 
The DAE derivatives that showed the first and second largest $\lambda_{\rm max}$ are CN-Me-CN-Me-H-Me and Me-CN-Me-CN-Me-H, whose structures are almost identical to each other since the unsubstituted DAE has $\it{C}_{2v}$-like symmetry. Those showing the third and fourth largest $\lambda_{\rm max}$ are Me-CN-Me-CN-Me-CN and CN-Me-CN-Me-CN-Me, whose structures are also almost identical to each other.
However, these four DAE derivatives, which correspond to the ground state and 1-3 excited states of the Ising Hamiltonian, may not be the suitable candidate for photopharmacology application due to their small values of Osc. 
The good balance between $\lambda_{\rm max}$ and Osc of Me-CN-Me-CN-H-H DAE derivative makes it a much better candidate as a photo switch molecule for photopharmacology application. 
These results suggest that the properties of $\lambda_{\rm max}$ and Osc can be tuned by various structural "handles".

As shown in Fig. \ref{fig:fig8}(b), since the HOMO and LUMO of the five candidate DAE derivatives are spatially fairly well separated, we can assume for simplicity that the absorption arises mainly from the transition between HOMO and LUMO configurations localized on the two thiophene rings, respectively. 
Thus, a large $\lambda_{\rm max}$ can be achieved by modifying the functional groups of the two thiophene rings moieties to increase their donor and acceptor strengths. 
However, the HOMO and LUMO (i.e. the donor and acceptor wavefunction) cannot be too well separated in space since Osc should remain sufficiently high to have a large absorption rate. 
We conclude that the donor and acceptor orbitals typically HOMO and LUMO should be somewhat, but not completely, spatially separated to obtain an ideal DAE candidate with large $\lambda_{\rm max}$ and Osc.

\section{Conclusions}

Integration of quantum chemistry simulations, machine learning techniques, and optimization calculations has been shown to be capable of accelerating the circle of novel materials discovery. 
However, even with these techniques, the task of molecular discovery for large systems is still daunting. 
In this work, we have introduced a combined quantum-classical method with the cVQD algorithm to calculate the excited states of classical Hamiltonian for efficiently yielding the targeted molecule candidates. 
This methodology is validated on searches for optimal photochromic DAE derivatives which have large $\lambda_{\rm max}$ and Osc for use as photo switches in photopharmacology applications. We have demonstrated that an FM machine learning model can accurately predict $\lambda_{\rm max}$ of 4096 DAE derivatives from a learning data set of 384 DAE derivative quantum chemistry results.       

With the Ising Hamiltonian constructed from the machine learning model, a 12-qubit cVQD calculation with a reference ground state obtained from VQE was performed on the the \texttt{statevector}, \texttt{qasm} simulator and quantum device of $ibm\_kawasaki$. Calculation results from the \texttt{statevector} simulator show that cVQD provides a robust approach to accurately calculate the excited states of Ising Hamiltonian. Moreover, we found excited state energies obtained from cVQD with limited shots on \texttt{qasm} simulator are within an average error of 0.038 to the exact values which outperforms conventional VQD by a factor of 2. 
The good performance of cVQD relies on the fact that the penalty overlap term in the cost function is almost accurately computed on classical computer.

Due to the influence of device noise, although the energies obtained from cVQD on $ibm\_kawasaki$ is on average 25\,nm higher than the calculation results on \texttt{statevector} simulator, the highest probability bitstings are the same. 
Thus, the results from $ibm\_kawasaki$ are also useful to find the optimal DAE derivatives. We also show that the energy difference can be improved to 3\,nm by mitigating the readout error and the decoherence error. 
More importantly, quantum chemistry calculations of five DAE derivatives candidates obtained from cVQD show how structural variation in the thiophene rings of DAE derivatives can tune the properties of $\lambda_{\rm max}$ and Osc and provide a path to design an ideal DAE candidate.

While calculations on much larger systems will be required to show the advantage of quantum computing, our work of searching optimal DAE candidates from 4096 DAE derivatives goes beyond proof of concept and paves the way for future work utilizing quantum computing to accelerate the discovery of novel materials. 
A possible next step of our study is to undertake the cVQD computations of large chemical space optimization on quantum devices with improved error rates, and utilizing more advanced error mitigation approaches.

\begin{acknowledgments}
Q.G., M.S., T.K., O.Y., H.T. and N.Y. acknowledge support from MEXT Quantum Leap Flagship Program Grant Number JPMXS0118067285 and JPMXS0120319794. N.Y. acknowledge support from JSPS KAKENHI Grant Number 20H05966. All the quantum chemistry calculations were performed on the NAYUTA Grid Cluster in Science \& Innovation Center of Mitsubishi Chemical Corporation. 
\end{acknowledgments}

\appendix
\section{IBM Quantum Kawasaki}\label{app:device}
\begin{figure}
\includegraphics[width=\columnwidth]{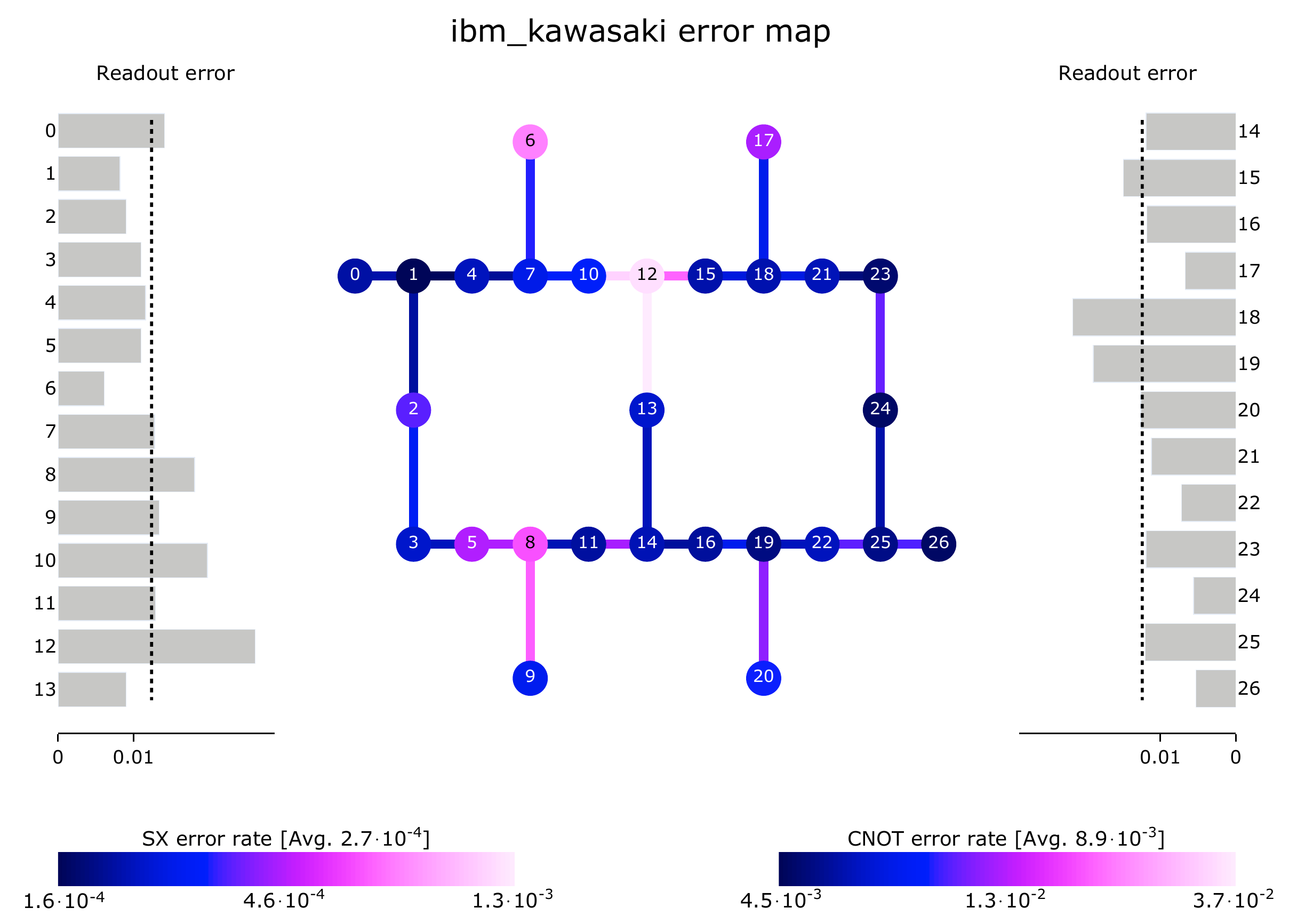}
\caption{\label{fig:fig11} Error rates of IBM Quantum Kawasaki on April 7th, 2023, as obtained from the device calibration data. 
The mean values of $T_{1}$ and $T_{2}$ (not shown) are $127~\rm{\mu s}$ and $132~\rm{\mu s}$, respectively.}
\end{figure}

The IBM Quantum Kawasaki device used in this work is a 27-qubit Falcon, revision 5.11, system. 
The entangling gate topology, based on CNOT gates, and error rates at the time of the experiments is shown in Fig.~\ref{fig:fig11}. In this work, we used qubits comprising the linear chain $[0,1,2,3,5,8,11,14,16,19,22,25]$.

\bibliographystyle{apsrev4-2}
\bibliography{design_dae}

\providecommand{\noopsort}[1]{}\providecommand{\singleletter}[1]{#1}%
\begin{thebibliography}{46}%
\makeatletter
\providecommand \@ifxundefined [1]{%
 \@ifx{#1\undefined}
}%
\providecommand \@ifnum [1]{%
 \ifnum #1\expandafter \@firstoftwo
 \else \expandafter \@secondoftwo
 \fi
}%
\providecommand \@ifx [1]{%
 \ifx #1\expandafter \@firstoftwo
 \else \expandafter \@secondoftwo
 \fi
}%
\providecommand \natexlab [1]{#1}%
\providecommand \enquote  [1]{``#1''}%
\providecommand \bibnamefont  [1]{#1}%
\providecommand \bibfnamefont [1]{#1}%
\providecommand \citenamefont [1]{#1}%
\providecommand \href@noop [0]{\@secondoftwo}%
\providecommand \href [0]{\begingroup \@sanitize@url \@href}%
\providecommand \@href[1]{\@@startlink{#1}\@@href}%
\providecommand \@@href[1]{\endgroup#1\@@endlink}%
\providecommand \@sanitize@url [0]{\catcode `\\12\catcode `\$12\catcode
  `\&12\catcode `\#12\catcode `\^12\catcode `\_12\catcode `\%12\relax}%
\providecommand \@@startlink[1]{}%
\providecommand \@@endlink[0]{}%
\providecommand \url  [0]{\begingroup\@sanitize@url \@url }%
\providecommand \@url [1]{\endgroup\@href {#1}{\urlprefix }}%
\providecommand \urlprefix  [0]{URL }%
\providecommand \Eprint [0]{\href }%
\providecommand \doibase [0]{https://doi.org/}%
\providecommand \selectlanguage [0]{\@gobble}%
\providecommand \bibinfo  [0]{\@secondoftwo}%
\providecommand \bibfield  [0]{\@secondoftwo}%
\providecommand \translation [1]{[#1]}%
\providecommand \BibitemOpen [0]{}%
\providecommand \bibitemStop [0]{}%
\providecommand \bibitemNoStop [0]{.\EOS\space}%
\providecommand \EOS [0]{\spacefactor3000\relax}%
\providecommand \BibitemShut  [1]{\csname bibitem#1\endcsname}%
\let\auto@bib@innerbib\@empty
\bibitem [{\citenamefont {Kyzer-Knapp}\ \emph {et~al.}(2015)\citenamefont
  {Kyzer-Knapp}, \citenamefont {Suh}, \citenamefont {Gomez-Bombarelli},
  \citenamefont {Aguilera-Iparraguirre},\ and\ \citenamefont
  {Aspuru-Guzik}}]{Pyzer-k_2015}%
  \BibitemOpen
  \bibfield  {author} {\bibinfo {author} {\bibfnamefont {E.~O.}\ \bibnamefont
  {Kyzer-Knapp}}, \bibinfo {author} {\bibfnamefont {C.}~\bibnamefont {Suh}},
  \bibinfo {author} {\bibfnamefont {R.}~\bibnamefont {Gomez-Bombarelli}},
  \bibinfo {author} {\bibfnamefont {J.}~\bibnamefont {Aguilera-Iparraguirre}},\
  and\ \bibinfo {author} {\bibfnamefont {A.}~\bibnamefont {Aspuru-Guzik}},\
  }\href@noop {} {\bibfield  {journal} {\bibinfo  {journal} {Annual Review of
  Materials Research}\ }\textbf {\bibinfo {volume} {45}},\ \bibinfo {pages}
  {195} (\bibinfo {year} {2015})}\BibitemShut {NoStop}%
\bibitem [{\citenamefont {Shoichet}(2004)}]{Shoichet-B_2004}%
  \BibitemOpen
  \bibfield  {author} {\bibinfo {author} {\bibfnamefont {B.~K.}\ \bibnamefont
  {Shoichet}},\ }\href@noop {} {\bibfield  {journal} {\bibinfo  {journal}
  {Nature}\ }\textbf {\bibinfo {volume} {432}},\ \bibinfo {pages} {862}
  (\bibinfo {year} {2004})}\BibitemShut {NoStop}%
\bibitem [{\citenamefont {Hachmann}\ \emph {et~al.}(2014)\citenamefont
  {Hachmann}, \citenamefont {Olivares-Amaya}, \citenamefont {Jinich},
  \citenamefont {Appleton}, \citenamefont {Blood-Forsythe}, \citenamefont
  {Seress}, \citenamefont {Román-Salgado}, \citenamefont {Trepte},
  \citenamefont {Atahan-Evrenk}, \citenamefont {Er}, \citenamefont {Shrestha},
  \citenamefont {Mondal}, \citenamefont {Sokolov}, \citenamefont {Bao},\ and\
  \citenamefont {Aspuru-Guzik}}]{Hachmann-J_2014}%
  \BibitemOpen
  \bibfield  {author} {\bibinfo {author} {\bibfnamefont {J.}~\bibnamefont
  {Hachmann}}, \bibinfo {author} {\bibfnamefont {R.}~\bibnamefont
  {Olivares-Amaya}}, \bibinfo {author} {\bibfnamefont {A.}~\bibnamefont
  {Jinich}}, \bibinfo {author} {\bibfnamefont {A.~L.}\ \bibnamefont
  {Appleton}}, \bibinfo {author} {\bibfnamefont {M.~A.}\ \bibnamefont
  {Blood-Forsythe}}, \bibinfo {author} {\bibfnamefont {L.~R.}\ \bibnamefont
  {Seress}}, \bibinfo {author} {\bibfnamefont {C.}~\bibnamefont
  {Román-Salgado}}, \bibinfo {author} {\bibfnamefont {K.}~\bibnamefont
  {Trepte}}, \bibinfo {author} {\bibfnamefont {S.}~\bibnamefont
  {Atahan-Evrenk}}, \bibinfo {author} {\bibfnamefont {S.}~\bibnamefont {Er}},
  \bibinfo {author} {\bibfnamefont {S.}~\bibnamefont {Shrestha}}, \bibinfo
  {author} {\bibfnamefont {R.}~\bibnamefont {Mondal}}, \bibinfo {author}
  {\bibfnamefont {A.}~\bibnamefont {Sokolov}}, \bibinfo {author} {\bibfnamefont
  {Z.}~\bibnamefont {Bao}},\ and\ \bibinfo {author} {\bibfnamefont
  {A.}~\bibnamefont {Aspuru-Guzik}},\ }\href@noop {} {\bibfield  {journal}
  {\bibinfo  {journal} {Energy \& Environmental Science}\ }\textbf {\bibinfo
  {volume} {7}},\ \bibinfo {pages} {698} (\bibinfo {year} {2014})}\BibitemShut
  {NoStop}%
\bibitem [{\citenamefont {Saal}\ \emph {et~al.}(2020)\citenamefont {Saal},
  \citenamefont {Oliynyk},\ and\ \citenamefont {Meredig}}]{Saal-J_2020}%
  \BibitemOpen
  \bibfield  {author} {\bibinfo {author} {\bibfnamefont {J.~E.}\ \bibnamefont
  {Saal}}, \bibinfo {author} {\bibfnamefont {A.~O.}\ \bibnamefont {Oliynyk}},\
  and\ \bibinfo {author} {\bibfnamefont {B.}~\bibnamefont {Meredig}},\
  }\href@noop {} {\bibfield  {journal} {\bibinfo  {journal} {Annual Review of
  Materials Research}\ }\textbf {\bibinfo {volume} {50}},\ \bibinfo {pages}
  {49} (\bibinfo {year} {2020})}\BibitemShut {NoStop}%
\bibitem [{\citenamefont {Ramakrishnan}\ \emph {et~al.}(2014)\citenamefont
  {Ramakrishnan}, \citenamefont {Dral}, \citenamefont {Rupp},\ and\
  \citenamefont {von Lilienfeld}}]{Ramakrishnan-R_2014}%
  \BibitemOpen
  \bibfield  {author} {\bibinfo {author} {\bibfnamefont {R.}~\bibnamefont
  {Ramakrishnan}}, \bibinfo {author} {\bibfnamefont {P.~O.}\ \bibnamefont
  {Dral}}, \bibinfo {author} {\bibfnamefont {M.}~\bibnamefont {Rupp}},\ and\
  \bibinfo {author} {\bibfnamefont {O.~A.}\ \bibnamefont {von Lilienfeld}},\
  }\href@noop {} {\bibfield  {journal} {\bibinfo  {journal} {Scientific Data}\
  }\textbf {\bibinfo {volume} {1}},\ \bibinfo {pages} {140022} (\bibinfo {year}
  {2014})}\BibitemShut {NoStop}%
\bibitem [{\citenamefont {Glavatskikh}\ \emph {et~al.}(2019)\citenamefont
  {Glavatskikh}, \citenamefont {Leguy}, \citenamefont {Hunault}, \citenamefont
  {Cauchy},\ and\ \citenamefont {Da~Mota}}]{Glavatskikh-M_2019}%
  \BibitemOpen
  \bibfield  {author} {\bibinfo {author} {\bibfnamefont {M.}~\bibnamefont
  {Glavatskikh}}, \bibinfo {author} {\bibfnamefont {J.}~\bibnamefont {Leguy}},
  \bibinfo {author} {\bibfnamefont {G.}~\bibnamefont {Hunault}}, \bibinfo
  {author} {\bibfnamefont {T.}~\bibnamefont {Cauchy}},\ and\ \bibinfo {author}
  {\bibfnamefont {B.}~\bibnamefont {Da~Mota}},\ }\href@noop {} {\bibfield
  {journal} {\bibinfo  {journal} {Journal of Cheminformatics}\ }\textbf
  {\bibinfo {volume} {11}},\ \bibinfo {pages} {69} (\bibinfo {year}
  {2019})}\BibitemShut {NoStop}%
\bibitem [{\citenamefont {Polishchuk}\ \emph {et~al.}(2013)\citenamefont
  {Polishchuk}, \citenamefont {Madzhidov},\ and\ \citenamefont
  {Varnek}}]{Polishchuk-P_2013}%
  \BibitemOpen
  \bibfield  {author} {\bibinfo {author} {\bibfnamefont {P.~G.}\ \bibnamefont
  {Polishchuk}}, \bibinfo {author} {\bibfnamefont {T.~I.}\ \bibnamefont
  {Madzhidov}},\ and\ \bibinfo {author} {\bibfnamefont {A.}~\bibnamefont
  {Varnek}},\ }\href@noop {} {\bibfield  {journal} {\bibinfo  {journal}
  {Journal of Computer-Aided Molecular Design}\ }\textbf {\bibinfo {volume}
  {27}},\ \bibinfo {pages} {675} (\bibinfo {year} {2013})}\BibitemShut
  {NoStop}%
\bibitem [{\citenamefont {Kirkpatrick}\ \emph {et~al.}(1983)\citenamefont
  {Kirkpatrick}, \citenamefont {Gelatt},\ and\ \citenamefont
  {Vecchi}}]{Kirkpatrick-S_1983}%
  \BibitemOpen
  \bibfield  {author} {\bibinfo {author} {\bibfnamefont {S.}~\bibnamefont
  {Kirkpatrick}}, \bibinfo {author} {\bibfnamefont {C.~D.~J.}\ \bibnamefont
  {Gelatt}},\ and\ \bibinfo {author} {\bibfnamefont {M.~P.}\ \bibnamefont
  {Vecchi}},\ }\href@noop {} {\bibfield  {journal} {\bibinfo  {journal}
  {Science}\ }\textbf {\bibinfo {volume} {220}},\ \bibinfo {pages} {671}
  (\bibinfo {year} {1983})}\BibitemShut {NoStop}%
\bibitem [{\citenamefont {Mitchell}(1998)}]{Mitchell-M_1998}%
  \BibitemOpen
  \bibfield  {author} {\bibinfo {author} {\bibfnamefont {M.}~\bibnamefont
  {Mitchell}},\ }\href@noop {} {\bibfield  {journal} {\bibinfo  {journal} {The
  MIT Press}\ } (\bibinfo {year} {1998})}\BibitemShut {NoStop}%
\bibitem [{\citenamefont {Gómez-Bombarelli†}\ \emph
  {et~al.}(2018)\citenamefont {Gómez-Bombarelli†}, \citenamefont {Wei},
  \citenamefont {Duvenaud}, \citenamefont {Hernandez-Lobato}, \citenamefont
  {Sanchez-Lengeling}, \citenamefont {Sheberla}, \citenamefont
  {Aguilera-Iparraguirre†}, \citenamefont {Hirzel}, \citenamefont {Adams},\
  and\ \citenamefont {Aspuru-Guzik}}]{Bombarelli_2018}%
  \BibitemOpen
  \bibfield  {author} {\bibinfo {author} {\bibfnamefont {R.}~\bibnamefont
  {Gómez-Bombarelli†}}, \bibinfo {author} {\bibfnamefont {J.~N.}\
  \bibnamefont {Wei}}, \bibinfo {author} {\bibfnamefont {D.}~\bibnamefont
  {Duvenaud}}, \bibinfo {author} {\bibfnamefont {J.~M.}\ \bibnamefont
  {Hernandez-Lobato}}, \bibinfo {author} {\bibfnamefont {B.}~\bibnamefont
  {Sanchez-Lengeling}}, \bibinfo {author} {\bibfnamefont {D.}~\bibnamefont
  {Sheberla}}, \bibinfo {author} {\bibfnamefont {J.}~\bibnamefont
  {Aguilera-Iparraguirre†}}, \bibinfo {author} {\bibfnamefont {T.~D.}\
  \bibnamefont {Hirzel}}, \bibinfo {author} {\bibfnamefont {R.~P.}\
  \bibnamefont {Adams}},\ and\ \bibinfo {author} {\bibfnamefont
  {A.}~\bibnamefont {Aspuru-Guzik}},\ }\href@noop {} {\bibfield  {journal}
  {\bibinfo  {journal} {ACS Cent. Sci.}\ }\textbf {\bibinfo {volume} {4}},\
  \bibinfo {pages} {268} (\bibinfo {year} {2018})}\BibitemShut {NoStop}%
\bibitem [{\citenamefont {Feynman}(1985)}]{Feynman-R_1985}%
  \BibitemOpen
  \bibfield  {author} {\bibinfo {author} {\bibfnamefont {R.~P.}\ \bibnamefont
  {Feynman}},\ }\href@noop {} {\bibfield  {journal} {\bibinfo  {journal}
  {Optics News}\ }\textbf {\bibinfo {volume} {11}},\ \bibinfo {pages} {11}
  (\bibinfo {year} {1985})}\BibitemShut {NoStop}%
\bibitem [{\citenamefont {Aspuru-Guzik}\ \emph {et~al.}(2005)\citenamefont
  {Aspuru-Guzik}, \citenamefont {Dutoi}, \citenamefont {Love},\ and\
  \citenamefont {Head-Gordon}}]{Aspuru-Guzik_2005}%
  \BibitemOpen
  \bibfield  {author} {\bibinfo {author} {\bibfnamefont {A.}~\bibnamefont
  {Aspuru-Guzik}}, \bibinfo {author} {\bibfnamefont {A.~D.}\ \bibnamefont
  {Dutoi}}, \bibinfo {author} {\bibfnamefont {P.~J.}\ \bibnamefont {Love}},\
  and\ \bibinfo {author} {\bibfnamefont {M.}~\bibnamefont {Head-Gordon}},\
  }\href@noop {} {\bibfield  {journal} {\bibinfo  {journal} {Science}\ }\textbf
  {\bibinfo {volume} {309}},\ \bibinfo {pages} {1704} (\bibinfo {year}
  {2005})}\BibitemShut {NoStop}%
\bibitem [{\citenamefont {Farhi}\ \emph {et~al.}(2014)\citenamefont {Farhi},
  \citenamefont {Goldstone},\ and\ \citenamefont {Gutmann}}]{farhi_2014}%
  \BibitemOpen
  \bibfield  {author} {\bibinfo {author} {\bibfnamefont {E.}~\bibnamefont
  {Farhi}}, \bibinfo {author} {\bibfnamefont {J.}~\bibnamefont {Goldstone}},\
  and\ \bibinfo {author} {\bibfnamefont {S.}~\bibnamefont {Gutmann}},\
  }\href@noop {} {\bibfield  {journal} {\bibinfo  {journal} {arXiv:1411.4028
  [quant-ph]}\ } (\bibinfo {year} {2014})}\BibitemShut {NoStop}%
\bibitem [{\citenamefont {Peruzzo}\ \emph {et~al.}(2014)\citenamefont
  {Peruzzo}, \citenamefont {McClean}, \citenamefont {Shadbolt}, \citenamefont
  {Yung}, \citenamefont {Zhou}, \citenamefont {Love}, \citenamefont
  {Aspuru-Guzik},\ and\ \citenamefont {O’Brien}}]{peruzzo_2014}%
  \BibitemOpen
  \bibfield  {author} {\bibinfo {author} {\bibfnamefont {A.}~\bibnamefont
  {Peruzzo}}, \bibinfo {author} {\bibfnamefont {J.}~\bibnamefont {McClean}},
  \bibinfo {author} {\bibfnamefont {P.}~\bibnamefont {Shadbolt}}, \bibinfo
  {author} {\bibfnamefont {M.-H.}\ \bibnamefont {Yung}}, \bibinfo {author}
  {\bibfnamefont {X.-Q.}\ \bibnamefont {Zhou}}, \bibinfo {author}
  {\bibfnamefont {P.~J.}\ \bibnamefont {Love}}, \bibinfo {author}
  {\bibfnamefont {A.}~\bibnamefont {Aspuru-Guzik}},\ and\ \bibinfo {author}
  {\bibfnamefont {J.~L.}\ \bibnamefont {O’Brien}},\ }\href@noop {} {\bibfield
   {journal} {\bibinfo  {journal} {Nat. Commun.}\ }\textbf {\bibinfo {volume}
  {5}},\ \bibinfo {pages} {4213} (\bibinfo {year} {2014})}\BibitemShut
  {NoStop}%
\bibitem [{\citenamefont {Kempe}\ \emph {et~al.}(2010)\citenamefont {Kempe},
  \citenamefont {Regev},\ and\ \citenamefont {Toner}}]{Kempe-J_2010}%
  \BibitemOpen
  \bibfield  {author} {\bibinfo {author} {\bibfnamefont {J.}~\bibnamefont
  {Kempe}}, \bibinfo {author} {\bibfnamefont {O.}~\bibnamefont {Regev}},\ and\
  \bibinfo {author} {\bibfnamefont {B.}~\bibnamefont {Toner}},\ }\href@noop {}
  {\bibfield  {journal} {\bibinfo  {journal} {SIAM Journal on Computing}\
  }\textbf {\bibinfo {volume} {39}},\ \bibinfo {pages} {3207} (\bibinfo {year}
  {2010})}\BibitemShut {NoStop}%
\bibitem [{\citenamefont {Egger}\ \emph {et~al.}(2021)\citenamefont {Egger},
  \citenamefont {Mareček},\ and\ \citenamefont {Woerner}}]{Egger-D_2021}%
  \BibitemOpen
  \bibfield  {author} {\bibinfo {author} {\bibfnamefont {D.~J.}\ \bibnamefont
  {Egger}}, \bibinfo {author} {\bibfnamefont {J.}~\bibnamefont {Mareček}},\
  and\ \bibinfo {author} {\bibfnamefont {S.}~\bibnamefont {Woerner}},\
  }\href@noop {} {\bibfield  {journal} {\bibinfo  {journal} {Quantum}\ }\textbf
  {\bibinfo {volume} {5}},\ \bibinfo {pages} {479} (\bibinfo {year}
  {2021})}\BibitemShut {NoStop}%
\bibitem [{\citenamefont {Gao}\ \emph {et~al.}(2021{\natexlab{a}})\citenamefont
  {Gao}, \citenamefont {Nakamura}, \citenamefont {Gujarati}, \citenamefont
  {Jones}, \citenamefont {Rice}, \citenamefont {Wood}, \citenamefont {Pistoia},
  \citenamefont {Garcia},\ and\ \citenamefont {Yamamoto}}]{Gao-1_2021}%
  \BibitemOpen
  \bibfield  {author} {\bibinfo {author} {\bibfnamefont {Q.}~\bibnamefont
  {Gao}}, \bibinfo {author} {\bibfnamefont {H.}~\bibnamefont {Nakamura}},
  \bibinfo {author} {\bibfnamefont {T.~P.}\ \bibnamefont {Gujarati}}, \bibinfo
  {author} {\bibfnamefont {G.~O.}\ \bibnamefont {Jones}}, \bibinfo {author}
  {\bibfnamefont {J.~E.}\ \bibnamefont {Rice}}, \bibinfo {author}
  {\bibfnamefont {S.~P.}\ \bibnamefont {Wood}}, \bibinfo {author}
  {\bibfnamefont {M.}~\bibnamefont {Pistoia}}, \bibinfo {author} {\bibfnamefont
  {J.~M.}\ \bibnamefont {Garcia}},\ and\ \bibinfo {author} {\bibfnamefont
  {N.}~\bibnamefont {Yamamoto}},\ }\href@noop {} {\bibfield  {journal}
  {\bibinfo  {journal} {The Journal of Physical Chemistry. A}\ }\textbf
  {\bibinfo {volume} {125}},\ \bibinfo {pages} {1827} (\bibinfo {year}
  {2021}{\natexlab{a}})}\BibitemShut {NoStop}%
\bibitem [{\citenamefont {Rice}\ \emph {et~al.}(2021)\citenamefont {Rice},
  \citenamefont {Gujarati}, \citenamefont {Motta}, \citenamefont {Takeshita},
  \citenamefont {Lee}, \citenamefont {Latone},\ and\ \citenamefont
  {Garcia}}]{Rice-J_2021}%
  \BibitemOpen
  \bibfield  {author} {\bibinfo {author} {\bibfnamefont {J.~E.}\ \bibnamefont
  {Rice}}, \bibinfo {author} {\bibfnamefont {T.~P.}\ \bibnamefont {Gujarati}},
  \bibinfo {author} {\bibfnamefont {M.}~\bibnamefont {Motta}}, \bibinfo
  {author} {\bibfnamefont {T.~Y.}\ \bibnamefont {Takeshita}}, \bibinfo {author}
  {\bibfnamefont {E.}~\bibnamefont {Lee}}, \bibinfo {author} {\bibfnamefont
  {J.~A.}\ \bibnamefont {Latone}},\ and\ \bibinfo {author} {\bibfnamefont
  {J.~M.}\ \bibnamefont {Garcia}},\ }\href@noop {} {\bibfield  {journal}
  {\bibinfo  {journal} {The Journal of Chemical Physics}\ }\textbf {\bibinfo
  {volume} {154}},\ \bibinfo {pages} {134115} (\bibinfo {year}
  {2021})}\BibitemShut {NoStop}%
\bibitem [{\citenamefont {Gao}\ \emph {et~al.}(2021{\natexlab{b}})\citenamefont
  {Gao}, \citenamefont {Jones}, \citenamefont {Motta}, \citenamefont
  {Sugawara}, \citenamefont {Watanabe}, \citenamefont {Kobayashi},
  \citenamefont {Watanabe}, \citenamefont {Ohnishi}, \citenamefont {Nakamura},\
  and\ \citenamefont {Yamamoto}}]{Gao_2021}%
  \BibitemOpen
  \bibfield  {author} {\bibinfo {author} {\bibfnamefont {Q.}~\bibnamefont
  {Gao}}, \bibinfo {author} {\bibfnamefont {G.~O.}\ \bibnamefont {Jones}},
  \bibinfo {author} {\bibfnamefont {M.}~\bibnamefont {Motta}}, \bibinfo
  {author} {\bibfnamefont {M.}~\bibnamefont {Sugawara}}, \bibinfo {author}
  {\bibfnamefont {H.~C.}\ \bibnamefont {Watanabe}}, \bibinfo {author}
  {\bibfnamefont {T.}~\bibnamefont {Kobayashi}}, \bibinfo {author}
  {\bibfnamefont {E.}~\bibnamefont {Watanabe}}, \bibinfo {author}
  {\bibfnamefont {Y.-y.}\ \bibnamefont {Ohnishi}}, \bibinfo {author}
  {\bibfnamefont {H.}~\bibnamefont {Nakamura}},\ and\ \bibinfo {author}
  {\bibfnamefont {N.}~\bibnamefont {Yamamoto}},\ }\href@noop {} {\bibfield
  {journal} {\bibinfo  {journal} {npj Computational Materials}\ }\textbf
  {\bibinfo {volume} {7}},\ \bibinfo {pages} {4213} (\bibinfo {year}
  {2021}{\natexlab{b}})}\BibitemShut {NoStop}%
\bibitem [{\citenamefont {Ibe}\ \emph {et~al.}(2021)\citenamefont {Ibe},
  \citenamefont {Nakagawa}, \citenamefont {Earnest}, \citenamefont {Yamamoto},
  \citenamefont {Mitarai}, \citenamefont {Gao},\ and\ \citenamefont
  {Kobayashi}}]{Ibe-Y_2021}%
  \BibitemOpen
  \bibfield  {author} {\bibinfo {author} {\bibfnamefont {Y.}~\bibnamefont
  {Ibe}}, \bibinfo {author} {\bibfnamefont {Y.~O.}\ \bibnamefont {Nakagawa}},
  \bibinfo {author} {\bibfnamefont {N.}~\bibnamefont {Earnest}}, \bibinfo
  {author} {\bibfnamefont {T.}~\bibnamefont {Yamamoto}}, \bibinfo {author}
  {\bibfnamefont {K.}~\bibnamefont {Mitarai}}, \bibinfo {author} {\bibfnamefont
  {Q.}~\bibnamefont {Gao}},\ and\ \bibinfo {author} {\bibfnamefont
  {T.}~\bibnamefont {Kobayashi}},\ }\href@noop {} {\bibfield  {journal}
  {\bibinfo  {journal} {Physical Review Research}\ }\textbf {\bibinfo {volume}
  {4}},\ \bibinfo {pages} {013173} (\bibinfo {year} {2021})}\BibitemShut
  {NoStop}%
\bibitem [{\citenamefont {Gao}\ \emph {et~al.}(2023)\citenamefont {Gao},
  \citenamefont {Jones}, \citenamefont {Sugawara}, \citenamefont {Kobayashi},
  \citenamefont {Yamashita}, \citenamefont {Kawaguchi}, \citenamefont
  {Tanaka},\ and\ \citenamefont {Yamamoto}}]{Gao-2_2023}%
  \BibitemOpen
  \bibfield  {author} {\bibinfo {author} {\bibfnamefont {Q.}~\bibnamefont
  {Gao}}, \bibinfo {author} {\bibfnamefont {G.~O.}\ \bibnamefont {Jones}},
  \bibinfo {author} {\bibfnamefont {M.}~\bibnamefont {Sugawara}}, \bibinfo
  {author} {\bibfnamefont {T.}~\bibnamefont {Kobayashi}}, \bibinfo {author}
  {\bibfnamefont {H.}~\bibnamefont {Yamashita}}, \bibinfo {author}
  {\bibfnamefont {H.}~\bibnamefont {Kawaguchi}}, \bibinfo {author}
  {\bibfnamefont {S.}~\bibnamefont {Tanaka}},\ and\ \bibinfo {author}
  {\bibfnamefont {N.}~\bibnamefont {Yamamoto}},\ }\href@noop {} {\bibfield
  {journal} {\bibinfo  {journal} {Intell. Comput.}\ }\textbf {\bibinfo {volume}
  {2}},\ \bibinfo {pages} {0037} (\bibinfo {year} {2023})}\BibitemShut
  {NoStop}%
\bibitem [{\citenamefont {Higgott}\ \emph {et~al.}(2019)\citenamefont
  {Higgott}, \citenamefont {Wang},\ and\ \citenamefont
  {Brierley}}]{Higgott-O_2019}%
  \BibitemOpen
  \bibfield  {author} {\bibinfo {author} {\bibfnamefont {O.}~\bibnamefont
  {Higgott}}, \bibinfo {author} {\bibfnamefont {D.}~\bibnamefont {Wang}},\ and\
  \bibinfo {author} {\bibfnamefont {S.}~\bibnamefont {Brierley}},\ }\href@noop
  {} {\bibfield  {journal} {\bibinfo  {journal} {Quantum}\ }\textbf {\bibinfo
  {volume} {3}},\ \bibinfo {pages} {156} (\bibinfo {year} {2019})}\BibitemShut
  {NoStop}%
\bibitem [{\citenamefont {Velema}\ \emph {et~al.}(2014)\citenamefont {Velema},
  \citenamefont {Szymanski},\ and\ \citenamefont {Feringa}}]{Velema-W_2014}%
  \BibitemOpen
  \bibfield  {author} {\bibinfo {author} {\bibfnamefont {W.~A.}\ \bibnamefont
  {Velema}}, \bibinfo {author} {\bibfnamefont {W.}~\bibnamefont {Szymanski}},\
  and\ \bibinfo {author} {\bibfnamefont {B.~L.}\ \bibnamefont {Feringa}},\
  }\href@noop {} {\bibfield  {journal} {\bibinfo  {journal} {JACS}\ }\textbf
  {\bibinfo {volume} {136}},\ \bibinfo {pages} {2178} (\bibinfo {year}
  {2014})}\BibitemShut {NoStop}%
\bibitem [{\citenamefont {Edwards}\ and\ \citenamefont
  {Aronson}(2000)}]{Edwards_2000}%
  \BibitemOpen
  \bibfield  {author} {\bibinfo {author} {\bibfnamefont {I.~R.}\ \bibnamefont
  {Edwards}}\ and\ \bibinfo {author} {\bibfnamefont {J.~K.}\ \bibnamefont
  {Aronson}},\ }\href@noop {} {\bibfield  {journal} {\bibinfo  {journal}
  {Lancet}\ }\textbf {\bibinfo {volume} {356}},\ \bibinfo {pages} {1255}
  (\bibinfo {year} {2000})}\BibitemShut {NoStop}%
\bibitem [{\citenamefont {Malhotra}\ and\ \citenamefont
  {Perry}(2003)}]{Malhotra_2003}%
  \BibitemOpen
  \bibfield  {author} {\bibinfo {author} {\bibfnamefont {V.}~\bibnamefont
  {Malhotra}}\ and\ \bibinfo {author} {\bibfnamefont {M.~C.}\ \bibnamefont
  {Perry}},\ }\href@noop {} {\bibfield  {journal} {\bibinfo  {journal} {Cancer
  Biol. Ther.}\ }\textbf {\bibinfo {volume} {2}},\ \bibinfo {pages} {S2}
  (\bibinfo {year} {2003})}\BibitemShut {NoStop}%
\bibitem [{\citenamefont {Carlet}\ \emph {et~al.}(2011)\citenamefont {Carlet},
  \citenamefont {Collignon}, \citenamefont {Goldmann}, \citenamefont
  {Goossens}, \citenamefont {Gyssens}, \citenamefont {Harbarth}, \citenamefont
  {Jarlier}, \citenamefont {Levy}, \citenamefont {N’Doye}, \citenamefont
  {Pittet}, \citenamefont {Seto}, \citenamefont {van~der Meer},\ and\
  \citenamefont {Voss}}]{Carlet_2011}%
  \BibitemOpen
  \bibfield  {author} {\bibinfo {author} {\bibfnamefont {J.}~\bibnamefont
  {Carlet}}, \bibinfo {author} {\bibfnamefont {P.}~\bibnamefont {Collignon}},
  \bibinfo {author} {\bibfnamefont {D.}~\bibnamefont {Goldmann}}, \bibinfo
  {author} {\bibfnamefont {H.}~\bibnamefont {Goossens}}, \bibinfo {author}
  {\bibfnamefont {I.~C.}\ \bibnamefont {Gyssens}}, \bibinfo {author}
  {\bibfnamefont {S.}~\bibnamefont {Harbarth}}, \bibinfo {author}
  {\bibfnamefont {V.}~\bibnamefont {Jarlier}}, \bibinfo {author} {\bibfnamefont
  {S.~B.}\ \bibnamefont {Levy}}, \bibinfo {author} {\bibfnamefont
  {B.}~\bibnamefont {N’Doye}}, \bibinfo {author} {\bibfnamefont
  {R.}~\bibnamefont {Pittet}, \bibfnamefont {D.~Richtmann}}, \bibinfo {author}
  {\bibfnamefont {W.~H.}\ \bibnamefont {Seto}}, \bibinfo {author}
  {\bibfnamefont {J.~W.~M.}\ \bibnamefont {van~der Meer}},\ and\ \bibinfo
  {author} {\bibfnamefont {A.}~\bibnamefont {Voss}},\ }\href@noop {} {\bibfield
   {journal} {\bibinfo  {journal} {Lancet}\ }\textbf {\bibinfo {volume}
  {378}},\ \bibinfo {pages} {369} (\bibinfo {year} {2011})}\BibitemShut
  {NoStop}%
\bibitem [{\citenamefont {Martinez}(2008)}]{Martinez_2008}%
  \BibitemOpen
  \bibfield  {author} {\bibinfo {author} {\bibfnamefont {J.~L.}\ \bibnamefont
  {Martinez}},\ }\href@noop {} {\bibfield  {journal} {\bibinfo  {journal}
  {Science}\ }\textbf {\bibinfo {volume} {321}},\ \bibinfo {pages} {365}
  (\bibinfo {year} {2008})}\BibitemShut {NoStop}%
\bibitem [{\citenamefont {Kudernac}\ \emph {et~al.}(2013)\citenamefont
  {Kudernac}, \citenamefont {Kobayashi}, \citenamefont {Uyama}, \citenamefont
  {Uchida}, \citenamefont {Nakamura},\ and\ \citenamefont
  {Feringa}}]{Kudernac-T_2013}%
  \BibitemOpen
  \bibfield  {author} {\bibinfo {author} {\bibfnamefont {T.}~\bibnamefont
  {Kudernac}}, \bibinfo {author} {\bibfnamefont {T.}~\bibnamefont {Kobayashi}},
  \bibinfo {author} {\bibfnamefont {A.}~\bibnamefont {Uyama}}, \bibinfo
  {author} {\bibfnamefont {K.}~\bibnamefont {Uchida}}, \bibinfo {author}
  {\bibfnamefont {S.}~\bibnamefont {Nakamura}},\ and\ \bibinfo {author}
  {\bibfnamefont {B.~L.}\ \bibnamefont {Feringa}},\ }\href@noop {} {\bibfield
  {journal} {\bibinfo  {journal} {The Journal of Physical Chemistry. A}\
  }\textbf {\bibinfo {volume} {117}},\ \bibinfo {pages} {8222} (\bibinfo {year}
  {2013})}\BibitemShut {NoStop}%
\bibitem [{\citenamefont {Irie}\ \emph {et~al.}(2014)\citenamefont {Irie},
  \citenamefont {Fukaminato}, \citenamefont {Matsuda},\ and\ \citenamefont
  {Kobatake}}]{Irie-M_2014}%
  \BibitemOpen
  \bibfield  {author} {\bibinfo {author} {\bibfnamefont {M.}~\bibnamefont
  {Irie}}, \bibinfo {author} {\bibfnamefont {T.}~\bibnamefont {Fukaminato}},
  \bibinfo {author} {\bibfnamefont {K.}~\bibnamefont {Matsuda}},\ and\ \bibinfo
  {author} {\bibfnamefont {S.}~\bibnamefont {Kobatake}},\ }\href@noop {}
  {\bibfield  {journal} {\bibinfo  {journal} {Chemical Reviews}\ }\textbf
  {\bibinfo {volume} {114}},\ \bibinfo {pages} {12174} (\bibinfo {year}
  {2014})}\BibitemShut {NoStop}%
\bibitem [{\citenamefont {Landrum}(2010)}]{Landrum-G_2010}%
  \BibitemOpen
  \bibfield  {author} {\bibinfo {author} {\bibfnamefont {G.}~\bibnamefont
  {Landrum}},\ }\href@noop {} {\bibfield  {journal} {\bibinfo  {journal}
  {https://www.rdkit.org/}\ } (\bibinfo {year} {2010})}\BibitemShut {NoStop}%
\bibitem [{\citenamefont {Rappe}\ \emph {et~al.}(1992)\citenamefont {Rappe},
  \citenamefont {Casewit}, \citenamefont {Colwell}, \citenamefont
  {Goddard~III},\ and\ \citenamefont {Skiff}}]{Rappe-A_1992}%
  \BibitemOpen
  \bibfield  {author} {\bibinfo {author} {\bibfnamefont {A.~K.}\ \bibnamefont
  {Rappe}}, \bibinfo {author} {\bibfnamefont {C.~J.}\ \bibnamefont {Casewit}},
  \bibinfo {author} {\bibfnamefont {K.~S.}\ \bibnamefont {Colwell}}, \bibinfo
  {author} {\bibfnamefont {W.~A.}\ \bibnamefont {Goddard~III}},\ and\ \bibinfo
  {author} {\bibfnamefont {W.~M.}\ \bibnamefont {Skiff}},\ }\href@noop {}
  {\bibfield  {journal} {\bibinfo  {journal} {J. Am. Chem. Soc.}\ }\textbf
  {\bibinfo {volume} {114}},\ \bibinfo {pages} {10024} (\bibinfo {year}
  {1992})}\BibitemShut {NoStop}%
\bibitem [{\citenamefont {Chai}\ and\ \citenamefont
  {Head-Gordon}(2008)}]{Chai-J_2008}%
  \BibitemOpen
  \bibfield  {author} {\bibinfo {author} {\bibfnamefont {J.-D.}\ \bibnamefont
  {Chai}}\ and\ \bibinfo {author} {\bibfnamefont {M.}~\bibnamefont
  {Head-Gordon}},\ }\href@noop {} {\bibfield  {journal} {\bibinfo  {journal}
  {Physical Chemistry Chemical Physics: PCCP}\ }\textbf {\bibinfo {volume}
  {10}},\ \bibinfo {pages} {6615} (\bibinfo {year} {2008})}\BibitemShut
  {NoStop}%
\bibitem [{\citenamefont {Yanai}\ \emph {et~al.}(2004)\citenamefont {Yanai},
  \citenamefont {Tew},\ and\ \citenamefont {Handy}}]{yanai_2004}%
  \BibitemOpen
  \bibfield  {author} {\bibinfo {author} {\bibfnamefont {T.}~\bibnamefont
  {Yanai}}, \bibinfo {author} {\bibfnamefont {D.~P.}\ \bibnamefont {Tew}},\
  and\ \bibinfo {author} {\bibfnamefont {N.~C.}\ \bibnamefont {Handy}},\
  }\href@noop {} {\bibfield  {journal} {\bibinfo  {journal} {Chem. Phys.
  Lett.}\ }\textbf {\bibinfo {volume} {393}},\ \bibinfo {pages} {51} (\bibinfo
  {year} {2004})}\BibitemShut {NoStop}%
\bibitem [{\citenamefont {Frisch}\ \emph {et~al.}(2009)\citenamefont {Frisch},
  \citenamefont {Trucks}, \citenamefont {Schlegel}, \citenamefont {Scuseria},
  \citenamefont {Robb}, \citenamefont {Cheeseman}, \citenamefont {Scalmani},
  \citenamefont {Barone}, \citenamefont {Nakatsuji}, \citenamefont {Li},
  \citenamefont {Caricato}, \citenamefont {Marenich}, \citenamefont {Bloino},
  \citenamefont {Janesko}, \citenamefont {Gomperts}, \citenamefont {Mennucci},
  \citenamefont {Hratchian}, \citenamefont {Ortiz}, \citenamefont {Izmaylov},
  \citenamefont {Sonnenberg}, \citenamefont {Williams-Young}, \citenamefont
  {Ding}, \citenamefont {Lipparini}, \citenamefont {Egidi}, \citenamefont
  {Goings}, \citenamefont {Peng}, \citenamefont {Petrone}, \citenamefont
  {Henderson}, \citenamefont {Ranasinghe}, \citenamefont {Zakrzewski},
  \citenamefont {Gao}, \citenamefont {Rega}, \citenamefont {Zheng},
  \citenamefont {Liang}, \citenamefont {Hada}, \citenamefont {Ehara},
  \citenamefont {Toyota}, \citenamefont {Fukuda}, \citenamefont {Hasegawa},
  \citenamefont {Ishida}, \citenamefont {Nakajima}, \citenamefont {Honda},
  \citenamefont {Kitao}, \citenamefont {Nakai}, \citenamefont {Vreven},
  \citenamefont {Throssell}, \citenamefont {Jr.}, \citenamefont {Peralta},
  \citenamefont {Ogliaro}, \citenamefont {Bearpark}, \citenamefont {Heyd},
  \citenamefont {Brothers}, \citenamefont {Kudin}, \citenamefont {Staroverov},
  \citenamefont {Keith}, \citenamefont {Kobayashi}, \citenamefont {Normand},
  \citenamefont {Raghavachari}, \citenamefont {Rendell}, \citenamefont
  {Burant}, \citenamefont {Iyengar}, \citenamefont {Tomasi}, \citenamefont
  {Cossi}, \citenamefont {Millam}, \citenamefont {Klene}, \citenamefont
  {Adamo}, \citenamefont {Cammi}, \citenamefont {Ochterski}, \citenamefont
  {Martin}, \citenamefont {Morokuma}, \citenamefont {Farkas}, \citenamefont
  {Foresman},\ and\ \citenamefont {Fox}}]{frisch_2009}%
  \BibitemOpen
  \bibfield  {author} {\bibinfo {author} {\bibfnamefont {M.~J.}\ \bibnamefont
  {Frisch}}, \bibinfo {author} {\bibfnamefont {G.~W.}\ \bibnamefont {Trucks}},
  \bibinfo {author} {\bibfnamefont {H.~B.}\ \bibnamefont {Schlegel}}, \bibinfo
  {author} {\bibfnamefont {G.~E.}\ \bibnamefont {Scuseria}}, \bibinfo {author}
  {\bibfnamefont {M.~A.}\ \bibnamefont {Robb}}, \bibinfo {author}
  {\bibfnamefont {J.~R.}\ \bibnamefont {Cheeseman}}, \bibinfo {author}
  {\bibfnamefont {G.}~\bibnamefont {Scalmani}}, \bibinfo {author}
  {\bibfnamefont {V.}~\bibnamefont {Barone}}, \bibinfo {author} {\bibfnamefont
  {H.}~\bibnamefont {Nakatsuji}}, \bibinfo {author} {\bibfnamefont
  {X.}~\bibnamefont {Li}}, \bibinfo {author} {\bibfnamefont {M.}~\bibnamefont
  {Caricato}}, \bibinfo {author} {\bibfnamefont {A.}~\bibnamefont {Marenich}},
  \bibinfo {author} {\bibfnamefont {J.}~\bibnamefont {Bloino}}, \bibinfo
  {author} {\bibfnamefont {B.~G.}\ \bibnamefont {Janesko}}, \bibinfo {author}
  {\bibfnamefont {R.}~\bibnamefont {Gomperts}}, \bibinfo {author}
  {\bibfnamefont {B.}~\bibnamefont {Mennucci}}, \bibinfo {author}
  {\bibfnamefont {H.~P.}\ \bibnamefont {Hratchian}}, \bibinfo {author}
  {\bibfnamefont {J.~V.}\ \bibnamefont {Ortiz}}, \bibinfo {author}
  {\bibfnamefont {A.~F.}\ \bibnamefont {Izmaylov}}, \bibinfo {author}
  {\bibfnamefont {J.~L.}\ \bibnamefont {Sonnenberg}}, \bibinfo {author}
  {\bibfnamefont {D.}~\bibnamefont {Williams-Young}}, \bibinfo {author}
  {\bibfnamefont {F.}~\bibnamefont {Ding}}, \bibinfo {author} {\bibfnamefont
  {F.}~\bibnamefont {Lipparini}}, \bibinfo {author} {\bibfnamefont
  {F.}~\bibnamefont {Egidi}}, \bibinfo {author} {\bibfnamefont
  {J.}~\bibnamefont {Goings}}, \bibinfo {author} {\bibfnamefont
  {B.}~\bibnamefont {Peng}}, \bibinfo {author} {\bibfnamefont {A.}~\bibnamefont
  {Petrone}}, \bibinfo {author} {\bibfnamefont {T.}~\bibnamefont {Henderson}},
  \bibinfo {author} {\bibfnamefont {D.}~\bibnamefont {Ranasinghe}}, \bibinfo
  {author} {\bibfnamefont {V.~G.}\ \bibnamefont {Zakrzewski}}, \bibinfo
  {author} {\bibfnamefont {J.}~\bibnamefont {Gao}}, \bibinfo {author}
  {\bibfnamefont {N.}~\bibnamefont {Rega}}, \bibinfo {author} {\bibfnamefont
  {G.}~\bibnamefont {Zheng}}, \bibinfo {author} {\bibfnamefont
  {W.}~\bibnamefont {Liang}}, \bibinfo {author} {\bibfnamefont
  {M.}~\bibnamefont {Hada}}, \bibinfo {author} {\bibfnamefont {M.}~\bibnamefont
  {Ehara}}, \bibinfo {author} {\bibfnamefont {K.}~\bibnamefont {Toyota}},
  \bibinfo {author} {\bibfnamefont {R.}~\bibnamefont {Fukuda}}, \bibinfo
  {author} {\bibfnamefont {J.}~\bibnamefont {Hasegawa}}, \bibinfo {author}
  {\bibfnamefont {M.}~\bibnamefont {Ishida}}, \bibinfo {author} {\bibfnamefont
  {T.}~\bibnamefont {Nakajima}}, \bibinfo {author} {\bibfnamefont
  {Y.}~\bibnamefont {Honda}}, \bibinfo {author} {\bibfnamefont
  {O.}~\bibnamefont {Kitao}}, \bibinfo {author} {\bibfnamefont
  {H.}~\bibnamefont {Nakai}}, \bibinfo {author} {\bibfnamefont
  {T.}~\bibnamefont {Vreven}}, \bibinfo {author} {\bibfnamefont
  {K.}~\bibnamefont {Throssell}}, \bibinfo {author} {\bibfnamefont {J.~A.~M.}\
  \bibnamefont {Jr.}}, \bibinfo {author} {\bibfnamefont {J.~E.}\ \bibnamefont
  {Peralta}}, \bibinfo {author} {\bibfnamefont {F.}~\bibnamefont {Ogliaro}},
  \bibinfo {author} {\bibfnamefont {M.}~\bibnamefont {Bearpark}}, \bibinfo
  {author} {\bibfnamefont {J.~J.}\ \bibnamefont {Heyd}}, \bibinfo {author}
  {\bibfnamefont {E.}~\bibnamefont {Brothers}}, \bibinfo {author}
  {\bibfnamefont {K.~N.}\ \bibnamefont {Kudin}}, \bibinfo {author}
  {\bibfnamefont {V.~N.}\ \bibnamefont {Staroverov}}, \bibinfo {author}
  {\bibfnamefont {T.}~\bibnamefont {Keith}}, \bibinfo {author} {\bibfnamefont
  {R.}~\bibnamefont {Kobayashi}}, \bibinfo {author} {\bibfnamefont
  {J.}~\bibnamefont {Normand}}, \bibinfo {author} {\bibfnamefont
  {K.}~\bibnamefont {Raghavachari}}, \bibinfo {author} {\bibfnamefont
  {A.}~\bibnamefont {Rendell}}, \bibinfo {author} {\bibfnamefont {J.~C.}\
  \bibnamefont {Burant}}, \bibinfo {author} {\bibfnamefont {S.~S.}\
  \bibnamefont {Iyengar}}, \bibinfo {author} {\bibfnamefont {J.}~\bibnamefont
  {Tomasi}}, \bibinfo {author} {\bibfnamefont {M.}~\bibnamefont {Cossi}},
  \bibinfo {author} {\bibfnamefont {J.~M.}\ \bibnamefont {Millam}}, \bibinfo
  {author} {\bibfnamefont {M.}~\bibnamefont {Klene}}, \bibinfo {author}
  {\bibfnamefont {C.}~\bibnamefont {Adamo}}, \bibinfo {author} {\bibfnamefont
  {R.}~\bibnamefont {Cammi}}, \bibinfo {author} {\bibfnamefont {J.~W.}\
  \bibnamefont {Ochterski}}, \bibinfo {author} {\bibfnamefont {R.~L.}\
  \bibnamefont {Martin}}, \bibinfo {author} {\bibfnamefont {K.}~\bibnamefont
  {Morokuma}}, \bibinfo {author} {\bibfnamefont {O.}~\bibnamefont {Farkas}},
  \bibinfo {author} {\bibfnamefont {J.~B.}\ \bibnamefont {Foresman}},\ and\
  \bibinfo {author} {\bibfnamefont {D.~J.}\ \bibnamefont {Fox}},\ }\href@noop
  {} {\bibinfo {title} {Gaussian 09, {Revision} {B}.01}} (\bibinfo {year}
  {2009})\BibitemShut {NoStop}%
\bibitem [{\citenamefont {Kochenberger}\ \emph {et~al.}(2014)\citenamefont
  {Kochenberger}, \citenamefont {Hao}, \citenamefont {Glover}, \citenamefont
  {Lewis}, \citenamefont {Lü}, \citenamefont {Wang},\ and\ \citenamefont
  {Wang}}]{kochenberger_2014}%
  \BibitemOpen
  \bibfield  {author} {\bibinfo {author} {\bibfnamefont {G.}~\bibnamefont
  {Kochenberger}}, \bibinfo {author} {\bibfnamefont {J.-K.}\ \bibnamefont
  {Hao}}, \bibinfo {author} {\bibfnamefont {F.}~\bibnamefont {Glover}},
  \bibinfo {author} {\bibfnamefont {M.}~\bibnamefont {Lewis}}, \bibinfo
  {author} {\bibfnamefont {Z.}~\bibnamefont {Lü}}, \bibinfo {author}
  {\bibfnamefont {H.}~\bibnamefont {Wang}},\ and\ \bibinfo {author}
  {\bibfnamefont {Y.}~\bibnamefont {Wang}},\ }\href@noop {} {\bibfield
  {journal} {\bibinfo  {journal} {J. Comb. Optim.}\ }\textbf {\bibinfo {volume}
  {28}},\ \bibinfo {pages} {58} (\bibinfo {year} {2014})}\BibitemShut {NoStop}%
\bibitem [{\citenamefont {Rendle}(2010)}]{rendle_fm_2010}%
  \BibitemOpen
  \bibfield  {author} {\bibinfo {author} {\bibfnamefont {S.}~\bibnamefont
  {Rendle}},\ }in\ \href@noop {} {\emph {\bibinfo {booktitle} {Proceedings of
  the 2010 {IEEE} {International} {Conference} on {Data} {Mining}}}},\ \bibinfo
  {series and number} {{ICDM} '10}\ (\bibinfo  {publisher} {IEEE Computer
  Society},\ \bibinfo {address} {USA},\ \bibinfo {year} {2010})\ pp.\ \bibinfo
  {pages} {995--1000}\BibitemShut {NoStop}%
\bibitem [{\citenamefont {Rendle}(2012)}]{rendle_2012}%
  \BibitemOpen
  \bibfield  {author} {\bibinfo {author} {\bibfnamefont {S.}~\bibnamefont
  {Rendle}},\ }\href@noop {} {\bibfield  {journal} {\bibinfo  {journal} {ACM
  Trans. Intell. Syst. Technol.}\ }\textbf {\bibinfo {volume} {3}},\ \bibinfo
  {pages} {57:1} (\bibinfo {year} {2012})}\BibitemShut {NoStop}%
\bibitem [{\citenamefont {Rendle}\ and\ \citenamefont
  {Schmidt-Thieme}(2010)}]{rendle_pairwise_2010}%
  \BibitemOpen
  \bibfield  {author} {\bibinfo {author} {\bibfnamefont {S.}~\bibnamefont
  {Rendle}}\ and\ \bibinfo {author} {\bibfnamefont {L.}~\bibnamefont
  {Schmidt-Thieme}},\ }in\ \href@noop {} {\emph {\bibinfo {booktitle}
  {Proceedings of the third {ACM} international conference on {Web} search and
  data mining - {WSDM} '10}}}\ (\bibinfo  {publisher} {ACM Press},\ \bibinfo
  {address} {New York, New York, USA},\ \bibinfo {year} {2010})\ pp.\ \bibinfo
  {pages} {81--90}\BibitemShut {NoStop}%
\bibitem [{\citenamefont {Ising}(1925)}]{ising_1925}%
  \BibitemOpen
  \bibfield  {author} {\bibinfo {author} {\bibfnamefont {E.}~\bibnamefont
  {Ising}},\ }\href@noop {} {\bibfield  {journal} {\bibinfo  {journal} {Z.
  Physik}\ }\textbf {\bibinfo {volume} {31}},\ \bibinfo {pages} {253} (\bibinfo
  {year} {1925})}\BibitemShut {NoStop}%
\bibitem [{\citenamefont {Tanaka}\ \emph {et~al.}(2017)\citenamefont {Tanaka},
  \citenamefont {Tamura},\ and\ \citenamefont {Chakrabarti}}]{tanaka-book}%
  \BibitemOpen
  \bibfield  {author} {\bibinfo {author} {\bibfnamefont {S.}~\bibnamefont
  {Tanaka}}, \bibinfo {author} {\bibfnamefont {R.}~\bibnamefont {Tamura}},\
  and\ \bibinfo {author} {\bibfnamefont {B.~K.}\ \bibnamefont {Chakrabarti}},\
  }\href@noop {} {\emph {\bibinfo {title} {Quantum Spin Glasses, Annealing and
  Computation}}}\ (\bibinfo  {publisher} {Cambridge University Press},\
  \bibinfo {year} {2017})\BibitemShut {NoStop}%
\bibitem [{\citenamefont {Tanahashi}\ \emph {et~al.}(2019)\citenamefont
  {Tanahashi}, \citenamefont {Takayanagi}, \citenamefont {Motohashi},\ and\
  \citenamefont {Tanaka}}]{tanahashi2019application}%
  \BibitemOpen
  \bibfield  {author} {\bibinfo {author} {\bibfnamefont {K.}~\bibnamefont
  {Tanahashi}}, \bibinfo {author} {\bibfnamefont {S.}~\bibnamefont
  {Takayanagi}}, \bibinfo {author} {\bibfnamefont {T.}~\bibnamefont
  {Motohashi}},\ and\ \bibinfo {author} {\bibfnamefont {S.}~\bibnamefont
  {Tanaka}},\ }\href {https://doi.org/10.7566/JPSJ.88.061010} {\bibfield
  {journal} {\bibinfo  {journal} {Journal of the Physical Society of Japan}\
  }\textbf {\bibinfo {volume} {88}},\ \bibinfo {pages} {061010} (\bibinfo
  {year} {2019})},\ \Eprint
  {https://arxiv.org/abs/https://doi.org/10.7566/JPSJ.88.061010}
  {https://doi.org/10.7566/JPSJ.88.061010} \BibitemShut {NoStop}%
\bibitem [{\citenamefont {Aleksandrowicz}\ \emph {et~al.}(2019)\citenamefont
  {Aleksandrowicz}, \citenamefont {Alexander}, \citenamefont {Barkoutsos},
  \citenamefont {Bello}, \citenamefont {Ben-Haim}, \citenamefont {Bucher},
  \citenamefont {Cabrera-Hernández}, \citenamefont {Carballo-Franquis},
  \citenamefont {Chen}, \citenamefont {Chen}, \citenamefont {Chow},
  \citenamefont {Córcoles-Gonzales}, \citenamefont {Cross}, \citenamefont
  {Cross}, \citenamefont {Cruz-Benito}, \citenamefont {Culver}, \citenamefont
  {González}, \citenamefont {Torre}, \citenamefont {Ding}, \citenamefont
  {Dumitrescu}, \citenamefont {Duran}, \citenamefont {Eendebak}, \citenamefont
  {Everitt}, \citenamefont {Sertage}, \citenamefont {Frisch}, \citenamefont
  {Fuhrer}, \citenamefont {Gambetta}, \citenamefont {Gago}, \citenamefont
  {Gomez-Mosquera}, \citenamefont {Greenberg}, \citenamefont {Hamamura},
  \citenamefont {Havlicek}, \citenamefont {Hellmers}, \citenamefont {Łukasz
  Herok}, \citenamefont {Horii}, \citenamefont {Hu}, \citenamefont {Imamichi},
  \citenamefont {Itoko}, \citenamefont {Javadi-Abhari}, \citenamefont
  {Kanazawa}, \citenamefont {Karazeev}, \citenamefont {Krsulich}, \citenamefont
  {Liu}, \citenamefont {Luh}, \citenamefont {Maeng}, \citenamefont {Marques},
  \citenamefont {Martín-Fernández}, \citenamefont {McClure}, \citenamefont
  {McKay}, \citenamefont {Meesala}, \citenamefont {Mezzacapo}, \citenamefont
  {Moll}, \citenamefont {Rodríguez}, \citenamefont {Nannicini}, \citenamefont
  {Nation}, \citenamefont {Ollitrault}, \citenamefont {O'Riordan},
  \citenamefont {Paik}, \citenamefont {Pérez}, \citenamefont {Phan},
  \citenamefont {Pistoia}, \citenamefont {Prutyanov}, \citenamefont {Reuter},
  \citenamefont {Rice}, \citenamefont {Davila}, \citenamefont {Rudy},
  \citenamefont {Ryu}, \citenamefont {Sathaye}, \citenamefont {Schnabel},
  \citenamefont {Schoute}, \citenamefont {Setia}, \citenamefont {Shi},
  \citenamefont {Silva}, \citenamefont {Siraichi}, \citenamefont {Sivarajah},
  \citenamefont {Smolin}, \citenamefont {Soeken}, \citenamefont {Takahashi},
  \citenamefont {Tavernelli}, \citenamefont {Taylor}, \citenamefont {Taylour},
  \citenamefont {Trabing}, \citenamefont {Treinish}, \citenamefont {Turner},
  \citenamefont {Vogt-Lee}, \citenamefont {Vuillot}, \citenamefont {Wildstrom},
  \citenamefont {Wilson}, \citenamefont {Winston}, \citenamefont {Wood},
  \citenamefont {Wood}, \citenamefont {Wörner}, \citenamefont {Akhalwaya},\
  and\ \citenamefont {Zoufal}}]{gadi_aleksandrowicz_2019}%
  \BibitemOpen
  \bibfield  {author} {\bibinfo {author} {\bibfnamefont {G.}~\bibnamefont
  {Aleksandrowicz}}, \bibinfo {author} {\bibfnamefont {T.}~\bibnamefont
  {Alexander}}, \bibinfo {author} {\bibfnamefont {P.}~\bibnamefont
  {Barkoutsos}}, \bibinfo {author} {\bibfnamefont {L.}~\bibnamefont {Bello}},
  \bibinfo {author} {\bibfnamefont {Y.}~\bibnamefont {Ben-Haim}}, \bibinfo
  {author} {\bibfnamefont {D.}~\bibnamefont {Bucher}}, \bibinfo {author}
  {\bibfnamefont {F.~J.}\ \bibnamefont {Cabrera-Hernández}}, \bibinfo {author}
  {\bibfnamefont {J.}~\bibnamefont {Carballo-Franquis}}, \bibinfo {author}
  {\bibfnamefont {A.}~\bibnamefont {Chen}}, \bibinfo {author} {\bibfnamefont
  {C.-F.}\ \bibnamefont {Chen}}, \bibinfo {author} {\bibfnamefont {J.~M.}\
  \bibnamefont {Chow}}, \bibinfo {author} {\bibfnamefont {A.~D.}\ \bibnamefont
  {Córcoles-Gonzales}}, \bibinfo {author} {\bibfnamefont {A.~J.}\ \bibnamefont
  {Cross}}, \bibinfo {author} {\bibfnamefont {A.}~\bibnamefont {Cross}},
  \bibinfo {author} {\bibfnamefont {J.}~\bibnamefont {Cruz-Benito}}, \bibinfo
  {author} {\bibfnamefont {C.}~\bibnamefont {Culver}}, \bibinfo {author}
  {\bibfnamefont {S.~D. L.~P.}\ \bibnamefont {González}}, \bibinfo {author}
  {\bibfnamefont {E.~D.~L.}\ \bibnamefont {Torre}}, \bibinfo {author}
  {\bibfnamefont {D.}~\bibnamefont {Ding}}, \bibinfo {author} {\bibfnamefont
  {E.}~\bibnamefont {Dumitrescu}}, \bibinfo {author} {\bibfnamefont
  {I.}~\bibnamefont {Duran}}, \bibinfo {author} {\bibfnamefont
  {P.}~\bibnamefont {Eendebak}}, \bibinfo {author} {\bibfnamefont
  {M.}~\bibnamefont {Everitt}}, \bibinfo {author} {\bibfnamefont {I.~F.}\
  \bibnamefont {Sertage}}, \bibinfo {author} {\bibfnamefont {A.}~\bibnamefont
  {Frisch}}, \bibinfo {author} {\bibfnamefont {A.}~\bibnamefont {Fuhrer}},
  \bibinfo {author} {\bibfnamefont {J.}~\bibnamefont {Gambetta}}, \bibinfo
  {author} {\bibfnamefont {B.~G.}\ \bibnamefont {Gago}}, \bibinfo {author}
  {\bibfnamefont {J.}~\bibnamefont {Gomez-Mosquera}}, \bibinfo {author}
  {\bibfnamefont {D.}~\bibnamefont {Greenberg}}, \bibinfo {author}
  {\bibfnamefont {I.}~\bibnamefont {Hamamura}}, \bibinfo {author}
  {\bibfnamefont {V.}~\bibnamefont {Havlicek}}, \bibinfo {author}
  {\bibfnamefont {J.}~\bibnamefont {Hellmers}}, \bibinfo {author} {\bibnamefont
  {Łukasz Herok}}, \bibinfo {author} {\bibfnamefont {H.}~\bibnamefont
  {Horii}}, \bibinfo {author} {\bibfnamefont {S.}~\bibnamefont {Hu}}, \bibinfo
  {author} {\bibfnamefont {T.}~\bibnamefont {Imamichi}}, \bibinfo {author}
  {\bibfnamefont {T.}~\bibnamefont {Itoko}}, \bibinfo {author} {\bibfnamefont
  {A.}~\bibnamefont {Javadi-Abhari}}, \bibinfo {author} {\bibfnamefont
  {N.}~\bibnamefont {Kanazawa}}, \bibinfo {author} {\bibfnamefont
  {A.}~\bibnamefont {Karazeev}}, \bibinfo {author} {\bibfnamefont
  {K.}~\bibnamefont {Krsulich}}, \bibinfo {author} {\bibfnamefont
  {P.}~\bibnamefont {Liu}}, \bibinfo {author} {\bibfnamefont {Y.}~\bibnamefont
  {Luh}}, \bibinfo {author} {\bibfnamefont {Y.}~\bibnamefont {Maeng}}, \bibinfo
  {author} {\bibfnamefont {M.}~\bibnamefont {Marques}}, \bibinfo {author}
  {\bibfnamefont {F.~J.}\ \bibnamefont {Martín-Fernández}}, \bibinfo {author}
  {\bibfnamefont {D.~T.}\ \bibnamefont {McClure}}, \bibinfo {author}
  {\bibfnamefont {D.}~\bibnamefont {McKay}}, \bibinfo {author} {\bibfnamefont
  {S.}~\bibnamefont {Meesala}}, \bibinfo {author} {\bibfnamefont
  {A.}~\bibnamefont {Mezzacapo}}, \bibinfo {author} {\bibfnamefont
  {N.}~\bibnamefont {Moll}}, \bibinfo {author} {\bibfnamefont {D.~M.}\
  \bibnamefont {Rodríguez}}, \bibinfo {author} {\bibfnamefont
  {G.}~\bibnamefont {Nannicini}}, \bibinfo {author} {\bibfnamefont
  {P.}~\bibnamefont {Nation}}, \bibinfo {author} {\bibfnamefont
  {P.}~\bibnamefont {Ollitrault}}, \bibinfo {author} {\bibfnamefont {L.~J.}\
  \bibnamefont {O'Riordan}}, \bibinfo {author} {\bibfnamefont {H.}~\bibnamefont
  {Paik}}, \bibinfo {author} {\bibfnamefont {J.}~\bibnamefont {Pérez}},
  \bibinfo {author} {\bibfnamefont {A.}~\bibnamefont {Phan}}, \bibinfo {author}
  {\bibfnamefont {M.}~\bibnamefont {Pistoia}}, \bibinfo {author} {\bibfnamefont
  {V.}~\bibnamefont {Prutyanov}}, \bibinfo {author} {\bibfnamefont
  {M.}~\bibnamefont {Reuter}}, \bibinfo {author} {\bibfnamefont
  {J.}~\bibnamefont {Rice}}, \bibinfo {author} {\bibfnamefont {A.~R.}\
  \bibnamefont {Davila}}, \bibinfo {author} {\bibfnamefont {R.~H.~P.}\
  \bibnamefont {Rudy}}, \bibinfo {author} {\bibfnamefont {M.}~\bibnamefont
  {Ryu}}, \bibinfo {author} {\bibfnamefont {N.}~\bibnamefont {Sathaye}},
  \bibinfo {author} {\bibfnamefont {C.}~\bibnamefont {Schnabel}}, \bibinfo
  {author} {\bibfnamefont {E.}~\bibnamefont {Schoute}}, \bibinfo {author}
  {\bibfnamefont {K.}~\bibnamefont {Setia}}, \bibinfo {author} {\bibfnamefont
  {Y.}~\bibnamefont {Shi}}, \bibinfo {author} {\bibfnamefont {A.}~\bibnamefont
  {Silva}}, \bibinfo {author} {\bibfnamefont {Y.}~\bibnamefont {Siraichi}},
  \bibinfo {author} {\bibfnamefont {S.}~\bibnamefont {Sivarajah}}, \bibinfo
  {author} {\bibfnamefont {J.~A.}\ \bibnamefont {Smolin}}, \bibinfo {author}
  {\bibfnamefont {M.}~\bibnamefont {Soeken}}, \bibinfo {author} {\bibfnamefont
  {H.}~\bibnamefont {Takahashi}}, \bibinfo {author} {\bibfnamefont
  {I.}~\bibnamefont {Tavernelli}}, \bibinfo {author} {\bibfnamefont
  {C.}~\bibnamefont {Taylor}}, \bibinfo {author} {\bibfnamefont
  {P.}~\bibnamefont {Taylour}}, \bibinfo {author} {\bibfnamefont
  {K.}~\bibnamefont {Trabing}}, \bibinfo {author} {\bibfnamefont
  {M.}~\bibnamefont {Treinish}}, \bibinfo {author} {\bibfnamefont
  {W.}~\bibnamefont {Turner}}, \bibinfo {author} {\bibfnamefont
  {D.}~\bibnamefont {Vogt-Lee}}, \bibinfo {author} {\bibfnamefont
  {C.}~\bibnamefont {Vuillot}}, \bibinfo {author} {\bibfnamefont {J.~A.}\
  \bibnamefont {Wildstrom}}, \bibinfo {author} {\bibfnamefont {J.}~\bibnamefont
  {Wilson}}, \bibinfo {author} {\bibfnamefont {E.}~\bibnamefont {Winston}},
  \bibinfo {author} {\bibfnamefont {C.}~\bibnamefont {Wood}}, \bibinfo {author}
  {\bibfnamefont {S.}~\bibnamefont {Wood}}, \bibinfo {author} {\bibfnamefont
  {S.}~\bibnamefont {Wörner}}, \bibinfo {author} {\bibfnamefont {I.~Y.}\
  \bibnamefont {Akhalwaya}},\ and\ \bibinfo {author} {\bibfnamefont
  {C.}~\bibnamefont {Zoufal}},\ }\href@noop {} {\bibinfo {title} {Qiskit: {An}
  {Open}-source {Framework} for {Quantum} {Computing}}} (\bibinfo {year}
  {2019})\BibitemShut {NoStop}%
\bibitem [{Note1()}]{Note1}%
  \BibitemOpen
  \bibinfo {note} {Note that, although the controlled-z gates in Fig.~\ref
  {fig:fig3}(a) can be executed in parallel, we choose to execute them in
  serial, with idle times populated with dynamical decoupling sequences as it
  yields higher fidelity results. This is likely due to the always-on-coupling
  of cross-resonance based devices, where the run-in-parallel gate operation
  can have larger error rates compared to isolated execution.}\BibitemShut
  {Stop}%
\bibitem [{\citenamefont {Nation}\ \emph {et~al.}(2021)\citenamefont {Nation},
  \citenamefont {Kang}, \citenamefont {Sundaresan},\ and\ \citenamefont
  {Gambetta}}]{Nation_2021}%
  \BibitemOpen
  \bibfield  {author} {\bibinfo {author} {\bibfnamefont {P.~D.}\ \bibnamefont
  {Nation}}, \bibinfo {author} {\bibfnamefont {H.}~\bibnamefont {Kang}},
  \bibinfo {author} {\bibfnamefont {N.}~\bibnamefont {Sundaresan}},\ and\
  \bibinfo {author} {\bibfnamefont {J.~M.}\ \bibnamefont {Gambetta}},\
  }\href@noop {} {\bibfield  {journal} {\bibinfo  {journal} {PRX QUANTUM}\
  }\textbf {\bibinfo {volume} {2}},\ \bibinfo {pages} {040326} (\bibinfo {year}
  {2021})}\BibitemShut {NoStop}%
\bibitem [{\citenamefont {Souza}\ \emph {et~al.}(2012)\citenamefont {Souza},
  \citenamefont {Álvarez},\ and\ \citenamefont {Suter}}]{Souza_2012}%
  \BibitemOpen
  \bibfield  {author} {\bibinfo {author} {\bibfnamefont {A.~M.}\ \bibnamefont
  {Souza}}, \bibinfo {author} {\bibfnamefont {G.~A.}\ \bibnamefont
  {Álvarez}},\ and\ \bibinfo {author} {\bibfnamefont {D.}~\bibnamefont
  {Suter}},\ }\href@noop {} {\bibfield  {journal} {\bibinfo  {journal}
  {Philosophical Transactions of the Royal Society A: Mathematical, Physical
  and Engineering Sciences}\ }\textbf {\bibinfo {volume} {370}},\ \bibinfo
  {pages} {4748} (\bibinfo {year} {2012})}\BibitemShut {NoStop}%
\bibitem [{\citenamefont {Gill}\ and\ \citenamefont
  {Meiboom}(1958)}]{Gill_1958}%
  \BibitemOpen
  \bibfield  {author} {\bibinfo {author} {\bibfnamefont {D.}~\bibnamefont
  {Gill}}\ and\ \bibinfo {author} {\bibfnamefont {S.}~\bibnamefont {Meiboom}},\
  }\href@noop {} {\bibfield  {journal} {\bibinfo  {journal} {Review of
  scientific instruments}\ }\textbf {\bibinfo {volume} {29}},\ \bibinfo {pages}
  {688} (\bibinfo {year} {1958})}\BibitemShut {NoStop}%
\end{thebibliography}%

\end{document}